%% file: reconstructing_and_resampling.tex
\documentclass[fleqn, usenatbib]{rasti}

\usepackage{newtxtext, newtxmath}
\usepackage[T1]{fontenc}

\usepackage{graphicx}
\usepackage{amsmath}
\usepackage{acronym}
\usepackage[separate-uncertainty = true, multi-part-units=single]{siunitx}
\usepackage{amsfonts}
\usepackage{comment}
\usepackage{subcaption}
\usepackage[capitalize]{cleveref}
\usepackage{bm}
\usepackage{listings}
\usepackage{xcolor}
\usepackage{xspace}
\usepackage{siunitx}
\usepackage{multirow}

\input{content/acronyms}
\input{content/macros}

\title[Reconstructing and resampling]{Reconstructing and resampling: a guide to utilising posterior samples from gravitational wave observations}

\author[G.~Ashton]{
Gregory Ashton$^{1,2}$,\thanks{E-mail: gregory.ashton@ligo.org}
\\
$^{1}$Physics Department, Royal Holloway, University of London, Egham Hill, Egham, TW20 0EX, United Kingdom\\
$^{2}$Mathematical Sciences, University of Southampton, Southampton SO17 1BJ, United Kingdom\\
}

\pubyear{2025}

\begin{document}
\label{firstpage}
\pagerange{\pageref{firstpage}--\pageref{lastpage}}
\maketitle

\begin{abstract}
\input{content/0-abstract}
\end{abstract}


\input{content/1-introduction}

\input{content/2-reconstructing}
\input{content/3-resampling}
\input{content/4-examples}
\input{content/5-discussion}

\section*{Acknowledgements}
We are thankful to Charlie Hoy for help with the software packaging provided by \PESUMMARY, to Colm Talbot and Tamasz Baka for useful discussions that refined the description of the resampling methods, and Michael Williams for useful comments in reviewing this work.

This work is supported by the Science and Technology Facilities Council (STFC) grant UKRI2488.
The authors are grateful for the computational resources provided by the LIGO Laboratory and supported by the National Science Foundation Grants PHY-0757058 and PHY-0823459.
This material is based upon work supported by NSF's LIGO Laboratory which is a major facility fully funded by the National Science Foundation.

We utilise the \texttt{Numpy} \citep{harris2020array} and \texttt{Scipy} library \citep{2020SciPy-NMeth} for data processing and analysis, we also use the \texttt{Matplotlib} \citep{Hunter:2007} library for visualisation.

\section*{Data Availability Statement}
We provide a data release associated with this paper \citep{data_release}, which contains the \PYTHON programs to recreate the results.
This includes programs to reconstruct the likelihood and priors from \ac{LVK} public data releases, and programs to resample the distribution for the given examples.
We include a lightweight \PYTHON module which implements the various resampling algorithms discussed within that can be easily ported or modified to users' projects.
If using this program (with modification), we ask authors to reference this work.

\bibliographystyle{mnras}
\bibliography{content/bibliography} 

\appendix

\input{content/6-appendix}

\bsp	
\label{lastpage}
\end{document}

%% file: content/acronyms.tex
\usepackage{amstext}

\makeatletter
\@ifpackageloaded{cleveref}{
  \AtBeginDocument{
    \def\ltx@label#1{\cref@label{#1}}
    \def\label@in@display@noarg#1{\cref@old@label@in@display{#1}}
    \def\label@in@mmeasure@noarg#1{%
      \begingroup%
        \measuring@false%
        \cref@old@label@in@display{#1}
      \endgroup}%
  }
}{}
\makeatother

\protected\def\protectedacused{\acused}

\acrodef{LIGO}[LIGO]{Laser Interferometer Gravitational-Wave Observatory}
\acrodef{LHO}[LHO]{\ac{LIGO} Hanford observatory}
\acrodef{LLO}[LLO]{\ac{LIGO} Livingston observatory}
\acrodef{KAGRA}[KAGRA]{KAGRA}\acused{KAGRA}
\acrodef{iKAGRA}[iKAGRA]{initial-phase \ac{KAGRA}}
\acrodef{bKAGRA}[bKAGRA]{baseline-design \ac{KAGRA}}
\acrodef{GEO}[GEO]{GEO\,600 \ac{GW} detector}
\acrodef{aLIGO}{Advanced \ac{LIGO}}
\acrodef{A+}{Advanced+ \ac{LIGO}}
\acrodef{Asharp}[\ensuremath{\text{A}^\sharp}]{\ac{LIGO} \acs{Asharp}}
\acrodef{AdV}{Advanced \acl{Virgo}}
\acrodef{AdV+}{Advanced \acl{Virgo}+}
\acrodef{Virgo}{Virgo}\acused{Virgo}
\acrodef{VirgoNEXT}[Virgo\_nEXT]{Virgo\_nEXT}\acused{VirgoNEXT}

\acrodef{LSC}[LSC]{\acs{LIGO} Scientific Collaboration}
\acrodef{LV}[LV]{\acs{LIGO}--\acs{Virgo} Collaboration\protect\protectedacused{LVC}}
\acrodef{LVC}[LV]{\acs{LIGO}--\acs{Virgo} Collaboration\protect\protectedacused{LV}}
\acrodef{LVK}[LVK]{\acs{LIGO}--Virgo--KAGRA}
\acrodef{IGWN}[IGWN]{International \ac{GWH} Observatory Network}

\acrodef{O1}[O1]{first observing run}
\acrodef{O2}[O2]{second observing run}
\acrodef{O3}[O3]{third observing run}
\acrodef{O3a}[O3a]{first half of the third observing run}
\acrodef{O3b}[O3b]{second half of the third observing run}
\acrodef{O3GK}[O3GK]{observing run}
\acrodef{O4}[O4]{fourth observing run}
\acrodef{O4a}[O4a]{first part of the fourth observing run}
\acrodef{O4b}[O4b]{second part of the fourth observing run}
\acrodef{O4c}[O4c]{third part of the fourth observing run}
\acrodef{O5}[O5]{fifth observing run}

\acrodef{BH}[BH]{black hole}
\acrodef{BBH}[BBH]{binary black hole}
\acrodef{BNS}[BNS]{binary neutron star}
\acrodef{IMBH}[IMBH]{intermediate-mass black hole}
\acrodef{NS}[NS]{neutron star}
\acrodef{BHNS}[BHNS]{black hole--neutron star binaries}
\acrodef{NSBH}[NSBH]{neutron star--black hole binary}
\acrodefplural{NSBH}[NSBH]{neutron star--black hole binaries}
\acrodef{PBH}[PBH]{primordial \ac{BH}}
\acrodef{CBC}[CBC]{compact binary coalescence}

\acrodef{GW}[GW]{gravitational wave\protect\protectedacused{GWH}}
\acrodef{GWH}[GW]{gravitational-wave\protect\protectedacused{GW}}
\acrodef{IFO}[IFO]{interferometer}
\acrodef{SNR}[SNR]{signal-to-noise ratio}
\acrodef{FAR}[FAR]{false alarm rate}
\acrodef{IFAR}[IFAR]{inverse false alarm rate}
\acrodef{FAP}[FAP]{false alarm probability}
\acrodef{PSD}[PSD]{power spectral density}
\acrodef{GR}[GR]{general relativity}
\acrodef{NR}[NR]{numerical relativity}
\acrodef{PN}[PN]{post-Newtonian}
\acrodef{EOB}[EOB]{effective-one-body}
\acrodef{ROM}[ROM]{reduced-order model}
\acrodef{IMR}[IMR]{inspiral--merger--ringdown}
\acrodef{PDF}[PDF]{probability density function}
\acrodef{PE}[PE]{parameter estimation}
\acrodef{CL}[CL]{credible level}
\acrodef{EOS}[EoS]{equation of state}
\acrodef{KLD}[KLD]{Kullback--Leibler divergence}
\acrodef{JSD}[JSD]{Jensen--Shannon divergence}
\acrodef{GCN}[GCN]{general coordinates network}
\acrodef{GWTC}[GWTC]{Gravitational-Wave Transient Catalog}
\acrodef{GWOSC}[GWOSC]{Gravitational Wave Open Science Center}

\acrodef{CWB}[cWB]{coherent WaveBurst}
\acrodef{LAL}[LAL]{\ac{LIGO} algorithm library}

\acrodef{CHRoCC}{central heating radius of curvature correction}
\acrodef{NonSENS}{non-stationary estimation and noise subtraction}

\acrodef{PTA}{Pulsar Timing Array}

\acrodef{MCMC}{Markov chain Monte Carlo}
\acrodef{ESS}{effective sample size}

\acrodef{RS}{rejection sampling}
\acrodef{IS}{importance sampling}
\acrodef{PSIS}{Pareto-smoothed \ac{IS}}
\acrodef{PSRS}{Pareto-smoothed \ac{RS}}

\acrodef{PP}{probability probabilty}
\acrodef{ASD}{amplitude spectral density}
\acrodef{IID}{independent and identically distributed}
\acrodef{KDE}{kernel density estimate}

%% file: content/macros.tex
\newcommand{\soft}[1]{\texttt{#1}}

\newcommand{\LALINFERENCE}{\soft{LALInference}\xspace}
\newcommand{\BAYESWAVE}{\soft{BayesWave}\xspace}
\newcommand{\BILBY}{\soft{Bilby}\xspace}
\newcommand{\RIFT}{\soft{RIFT}\xspace}
\newcommand{\LALSUITE}{\soft{LALSuite}\xspace}
\newcommand{\BILBYPIPE}{\soft{BilbyPipe}\xspace}

\newcommand{\PESUMMARY}{\soft{PESummary}\xspace}
\newcommand{\PYTHON}{\soft{Python}\xspace}

\newcommand{\SCIPY}{\soft{SciPy}\xspace}

\newcommand{\DYNESTY}{\soft{Dynesty}\xspace}

\newcommand{\CONDA}{\soft{conda}\xspace}
\newcommand{\LALSIMULATION}{\soft{LALSimulation}\xspace}

\newcommand{\IMRPhenomXP}{\soft{IMRPhenomXP}\xspace}
\newcommand{\IMRPhenomXPHM}{\soft{IMRPhenomXPHM}\xspace}

\newcommand{\data}{\ensuremath{d}\xspace}

\newcommand{\fdata}{\ensuremath{\tilde{d}}\xspace}
\newcommand{\model}{\ensuremath{M}\xspace}
\newcommand{\parameters}{\ensuremath{\theta}\xspace}
\newcommand{\likelihood}{\ensuremath{\mathcal{L}}\xspace}
\newcommand{\prior}{\ensuremath{\pi}\xspace}
\newcommand{\evidence}{\ensuremath{\mathcal{Z}}\xspace}
\newcommand{\fsample}{\ensuremath{f_s}\xspace}

\newcommand{\BF}{\ensuremath{{\mathcal{B}}}}
\newcommand{\avew}{\ensuremath{{\langle w \rangle}}}
\newcommand{\Ness}{\ensuremath{N_{\rm ess}}}

\newcommand{\expect}[1]{\left\langle #1 \right\rangle}

%% file: content/0-abstract.tex
The LIGO, Virgo, and KAGRA (LVK) gravitational-wave observatories have opened new scientific research in astrophysics, fundamental physics, and cosmology.
The collaborations that build and operate these observatories release the interferometric strain data as well as a catalogue of observed signals with accompanying Bayesian posterior distributions.
These posteriors, in the form of equally-weighted samples, form a dataset that is used by a multitude of further analyses seeking to constrain the population of merging black holes, identify lensed pairs of signals, and much more.
However, many of these analyses rely, often implicitly, on the ability to reconstruct the likelihood and prior from the inputs to the analysis and apply resampling (a statistical technique to generate new samples varying the underlying analysis assumptions).
In this work, we first provide a guide on how to reconstruct and modify the posterior density accurately from the inputs for analyses performed with the \BILBY inference library.
We then demonstrate and compare resampling techniques to produce new posterior sample sets and discuss Pareto-smoothing to improve the efficiency.
Finally, we provide examples of how to use resampling to study observed gravitational-wave signals.
We hope this guide provides a useful resource for those wishing to use open data products from the LVK for gravitational-wave astronomy.

%% file: content/1-introduction.tex
\section{Introduction}
\label{eqn:introduction}

Gravitational wave observations of merging black holes and neutron stars are now routinely made by interferometric gravitational wave detectors.
There are currently four operational kilometre-scale interferometes: \aclu{LIGO} \citep[LIGO:][]{LIGOScientific:2014pky}, Virgo \citep{VIRGO:2014yos}, and KAGRA \citep{KAGRA:2020tym}.
Their raw output is a time series of calibrated strain, representing the relative difference in arm lengths for each instrument.
The \ac{LVK} Collaborations publish this data \citep{GWTC-4-data} in addition to the \acl{GWTC} \citep[\acsu{GWTC}:][]{GWTC-4-intro}, containing a list of candidate signals and measurements of their Bayesian posterior distributions \citep{GWTC-4-results}.

The posterior distribution \citep[see][for a discussion on why a Bayesian framework is used]{Thrane:2018qnx} is defined as the probability density of a set of source parameters $\parameters$ conditional on the data \data and model \model (which implicitly defines the parameters) and can be calculated up to a normalisation factor from
\begin{align}
p(\parameters | \data, \model) = \likelihood(\data| \theta, \model) \prior(\theta | \model)\,,
\label{eqn:posterior}
\end{align}
where $\likelihood(\data| \parameters, \model)$ is the likelihood\footnote{Formally, the likelihood is the conditional probability of the data conditional on the parameters, but in parameter estimation, where the observed data is fixed, we treat it as a function of the parameters.} and $\prior(\parameters | \model)$ is the prior probability distribution of the parameters conditioned only on the model.
The normalisation factor that ensures $\int p(\parameters | \data, \model) \, d\theta=1$ can also be calculated and is referred to as the evidence:
\begin{align}
\evidence(\data | \model) = \int \likelihood(\data| \theta, \model) \prior(\theta | \model) \, d\parameters\,.
\label{eqn:evidence}
\end{align}

The \ac{LVK} analysis methodology for approximating \cref{eqn:posterior} for observed signals is described in detail in Section 5 of \citet{GWTC-4-methods}.
In brief, for transient gravitational-wave data analysis, analysis is performed on a time series of consecutive strain data observations, which we denote by \data and define its duration to be $T$ and sampling frequency $\fsample$.
Inference is performed on the frequency series \fdata generated by applying a fast Fourier transform and dividing by \fsample (a normalisation choice; see \citet{Thrane:2018qnx}).
For signals lasting up to a few hundred seconds, we assume the non-astrophysical noise to be generated by a stationary additive coloured Gaussian noise process with a \ac{PSD} $P$ such that the likelihood for the $j$th frequency bin is given by
\begin{equation}
    \likelihood(\data_j | \parameters, \model) = 
    \frac{1}{2\pi P_j}
    \exp\left(
    -\frac{2}{T}\frac{|\fdata_j - \tilde{\mu}_j(\parameters)|^2}{P_j}
    \right)\,,
    \label{eqn:likelihood}
\end{equation}
where $\tilde{\mu}_j(\parameters)$ is the frequency-domain astrophysical signal model projected onto the response of the detector.
The full likelihood is then calculated from the product over the $j$ frequency bins, i.e.
\begin{equation}
    \likelihood(\data | \theta, \model) = \prod_{j} \likelihood(\data_j | \theta, \model)\,.
\end{equation} 
In practise, this is computed as the sum in log-space for numerical stability.

Typically, $\tilde{\mu}_j(\parameters)$ is a waveform approximant that models the inspiral--merger--ringdown of the \ac{CBC} signal utilising either phenomenological fits \citep{Ajith:2007qp}, the effective one-body approach \citep{Buonanno:1998gg,Buonanno:2000ef}, or numerical-relativity surrogates \citep{Blackman:2017dfb}.
To date, \BILBY uses the \LALSUITE \citep{lalsuite, Wette:2020air} library to access these waveforms.

Meanwhile, the prior distribution $\prior(\theta | \model)$ is either generally weakly astrophysically motivated or informed by the observed population of \ac{CBC} sources \citep{GWTC-4-methods}.

\cref{eqn:posterior} is not solvable in closed form for the \ac{CBC} likelihood and prior, and the distribution is typically non-Gaussian, multimodal, and contains strong curved degeneracies.
Therefore, computational Bayesian inference methods such as \acl{MCMC}~\citep[\acsu{MCMC};][]{Hastings:1970aa} and nested sampling~\citep{Skilling:2006gxv} are used with post-processing to produce a set of equally-weighted posterior samples $\theta_i\sim p(\theta | \data, \model)$ and an estimate of the evidence $\evidence(\data| \model)$.
To date, the \ac{LVK} has used \LALINFERENCE~\citep{Veitch:2014wba}, \RIFT~\citep{Pankow:2015cra,Lange:2017wki,Wysocki:2019grj}, and \BILBY~\citep{Ashton:2018jfp,Romero-Shaw:2020owr}, with an adaptation of the \DYNESTY~\citep{Speagle:2019ivv} nested sampling implementation, to produce posterior samples.
In this work, we will consider the \ac{LVK} posteriors produced by \BILBY and packaged using \PESUMMARY~\citep{Hoy:2020vys}.
This is motivated by the ease of use in reconstructing the posteriors using \BILBY (part of its design philosophy) and the wide applicability: \BILBY posteriors have been estimated for all signals above threshold for parameter estimation from the first observing run until the most recent release, see GWTC-2.1~\citep{LIGOScientific:2021usb}, GWTC-3.0~\citep{KAGRA:2021vkt}, and GWTC-4.0~\citep{GWTC-4-results}.

The posterior samples from a given observed signal form a data set that can be used for further analyses.
For example, taking samples from a set of highly probable signals, the samples form an input data set that can be used to constrain the astrophysical distribution of merging black holes \citep[see, e.g.][]{Talbot:2018cva, GWTC-4-population} or measure the cosmological properties of our universe \citep[see, e.g.][]{Schutz:1986gp, GWTC-4-cosmology}.
These methods typically perform inference on a hierarchical population model using the \ac{GWTC} posteriors as input data \citep{Thrane:2018qnx}.
On the other hand, considering individual events, studies can be conducted using the posterior samples to examine the inferences under alternative assumptions about the waveform model, prior distribution, or data.
Underlying all these methods is the technique known as \emph{resampling}, which utilises a set of weights, constructed from the ratio of the new to the old posterior density, to generate a new set of samples under some changed assumptions.
Resampling is an alternative to a re-analysis (i.e. applying stochastic sampling to estimate the posterior afresh), which can be computationally prohibitive.

Two types of resampling are in common use: \ac{RS} and \ac{IS}.
For \ac{RS} (discussed further in \cref{sec:rs}), samples are accepted into the new set in proportion to their weight.
Meanwhile, \ac{IS} (discussed further in \cref{sec:is}) is a broad term, often without a consensus definition, but generally refers to the process of using the weights to determine the importance of the samples.
In some use cases, this involves using the weights and samples together (e.g. to create a weighted histogram).
Alternatively, some use cases apply multinomial sampling to produce a new sample set according to the probabilities provided by the weights (either with or without replacement).
In this work, we refer to the former as \ac{IS} and the latter as multinomial-\ac{IS}.

Resampling techniques have long been used in the field.
For example, both \ac{MCMC} and nested sampling utilise resampling to produce sample sets;
\citet{Rover:2006ni} used multinomial-\ac{IS} to initialise a \ac{MCMC} sampler;
\citet{Payne:2019wmy} demonstrated the use of \ac{IS} to utilise higher-order gravitational-wave models;
\citet{Payne:2020myg} extended this to marginalise over the calibration model (and this was implemented in \citet{LIGOScientific:2021usb});
\citet{Baka2025} explored \ac{RS} to correct for calibration misspecification;
\citet{Tiwari:2017ndi} and \citet{Talbot:2019okv} used \ac{IS} to estimate the \ac{BBH} population sensitivity;
\ac{RS} is used to apply a cosmological prior to \ac{LVK} analyses \citep{GWTC-4-methods}, and an astrophysically-motivated prior
\citep{Chattopadhyay:2024hsf, GWTC-4-results};
\citet{Dax:2022pxd} uses \ac{IS} with a proposal distribution provided by neural posterior estimation (but generally the technique can be applied to any pre-computed sample set with a tractable posterior);
\citet{Williams:2025szm} investigated sequential-Monte Carlo approaches, an extension of \ac{IS};
and finally, \citet{Hourihane:2025vxc}, use \ac{IS} changing the data in the likelihood to study the impact of transient non-Gaussian noise.
There are many more examples besides this in the literature.
However, we caution that the terms are often confused and there are additional terms such as ``reweighting'' that we choose to avoid since this is used to describe both \ac{RS} and \ac{IS}.

The goal of this work is to provide a simple guide to the application of resampling, along with a discussion of reconstructing the likelihood and prior distribution.
We begin in \cref{sec:reconstructing} by discussing how the likelihood and prior can be calculated from the released posterior samples, including a comparison with the original calculations stored in data files.
We also provide estimates of the accuracy achievable given practical hardware and software considerations.
Then, in \cref{sec:resampling}, we go on to introduce and compare the methodology of resampling, including \ac{RS} and \ac{IS} and introduce Pareto-smoothing as a means to improve the efficiency.
In \cref{sec:example}, we provide examples using publicly available data sets and applying resampling to illustrate the possibilities. 
Finally, we conclude with a discussion in \cref{sec:discussion}.
Readers may find it useful to refer to the data release associated with this work \citep{data_release}, which contains worked examples.

%% file: content/2-reconstructing.tex
\section{Reconstructing the likelihood and prior}
\label{sec:reconstructing}

In this Section, we will detail how to reconstruct the likelihood and prior for samples drawn from the posterior distribution presented in the digital \ac{GWTC}.
Specifically, we consider existing analyses of gravitational-wave signals using the \BILBY inference library and then packaged using \PESUMMARY.
Examples of such files can be found as part of the data released with GWTC-4.0~\citep{GWTC-4.0:PE}.

Each file stores a set of posterior samples $\{\parameters_i\}$.
Additionally, users will find columns corresponding to the log-likelihood and log-prior (in all cases, these are natural logarithms).
We refer to these as the \emph{stored} values and denote them with a sub-script $S$ (e.g. $\ln\likelihood_S$ and $\ln\prior_S$) to differentiate them from the \emph{reconstructed} values that we denote with a sub-script $R$
(e.g. $\ln\likelihood_R$ and $\ln\prior_R$).
For accurate reconstruction of the log-likelihood and log-prior for a sample, it is important to match the original computation.
Any differences can manifest as either systematic shifts (which could entirely bias any subsequent resampling) or random shifts (which, if sufficiently small, may be negligible).
In the following, we address each point in turn.

\subsection{Computing environment}

\begin{table*}
    \centering
    \begin{tabular}{c|c|c|c|c|c|c}
         Analysis & \CONDA environment & \PYTHON & \BILBY & \BILBYPIPE & \DYNESTY & \LALSIMULATION \\ \hline\
         GWTC-2.1 & \soft{igwn-py38-20210107} & 3.8 & 1.0.4 & 1.0.2 & 1.0.1 & 2.4.0 \\
         GWTC-3.0 & \soft{igwn-py38-20210107} & 3.8 & 1.0.4 & 1.0.2 & 1.0.1 & 2.4.0 \\
         GWTC-4.0 & \soft{igwn-py310-20241106} & 3.10 & 2.2.2.1 & 1.4.0 & 2.1.4 & 6.0.0\\
    \end{tabular}
    \caption{A table detailing the \CONDA environment and specific top-level packages used in recent \ac{GWTC} updates. Full details of the environments can be found in the data release associated with this work.}
    \label{tab:environments}
\end{table*}

There are two crucial aspects to consider regarding the computing environment used to reconstruct the likelihood and prior. 

First, the versions of the software should match those used in the analysis: this is because changes to the software may reflect changes to the underlying assumptions.
For example, if the default assumptions about data processing change between versions of the software, this will induce systematic shifts in the likelihood.
Similarly, if the waveform model is not identical (including specific options hardcoded into the environment), this can also induce systematic and random shifts in the likelihood.
Information about the software packages used in the analysis is packaged as part of the \PESUMMARY results files (see the data release for details on how to access the information).

However, for all \ac{LVK} analyses, the \CONDA (\href{https://docs.anaconda.com/}{docs.anaconda.com/}) package manager has been used to fully specify the environment, and typically it is easier to use a \CONDA environment rather than attempt to match packages individually.
Therefore, in \cref{tab:environments}, we list the \CONDA environments used for pertinent updates to the GWTC alongside the versions of the packages that have the most significant impact (namely the sampling and simulation libraries); full details of the environments can be found in the data release of this work.

Second, for exact reconstruction, the hardware should also be identical.
This is because the underlying hardware determines the implementation of floating-point arithmetic: differences in this implementation will change the byte-level results.
There may also be platform-specific optimisation of linear algebra libraries that can differ even if the hardware is identical.
Comparing an analysis run on an Intel Xeon Processor (Skylake) and then reconstructed on an Apple M3 Pro, we find the distributions of log-likelihoods are centred on zero but with a standard deviation of $\approx3\times10^{-6}$.
Meanwhile, when they are reconstructed on identical hardware, we find the log-likelihoods can be reconstructed up to floating-point precision.
However, typically, it is not possible to match the hardware used in the analysis.
But, as we will show later, differences of less than 1 part in a million are sufficiently small that they have a negligible impact on the capacity to resample the posterior.

\subsection{Likelihood}

To reconstruct the likelihood, \cref{eqn:likelihood}, for a given sample, we must match the data, \ac{PSD}, waveform model, and likelihood configuration.
In the data release, we demonstrate how this can be achieved, taking the settings from the configuration file packaged as part of the data release.
In the following, we provide a discussion of specific elements.

\emph{The likelihood ratio} -- \cref{eqn:likelihood} is the likelihood of the data under the model comprised of an additive signal and noise.
However, we can see that if $\mu_j=0\; \forall\; j$, i.e. there is no signal, just noise, then we can also introduce $\likelihood(d | N)$, the noise likelihood.
Since the noise likelihood is $\theta$-independent, it is common during sampling to use the likelihood ratio
\begin{equation}
    \Lambda(d | \theta, \model) = \frac{\likelihood(d | \theta, \model)}{\likelihood(d | N)}\,,
    \label{eqn:likelihood-ratio}
\end{equation}
rather than the likelihood itself.
While this choice does not change the parameter estimates, it is often used because the evidence estimate obtained (e.g. from nested sampling) is the signal vs noise Bayes factor, i.e.
$\BF = \evidence(d| \model)/\evidence(d| N)$,
which is easier to interpret as a diagnostic during sampling rather than $\evidence{(d|\model)}$ itself.

When reconstructing the likelihood, it is important to distinguish between the likelihood and the likelihood ratio.
For resampling, one only needs to be consistent
However, evidence estimates computed from the likelihood will yield measurements of the signal evidence.
In contrast, those computed from the likelihood ratio will yield measurements of the signal vs noise Bayes factor.
\BILBY can compute either as demonstrated in the data release.

\emph{Data} -- The raw strain data used in \ac{LVK} analyses is available from the \ac{GWOSC} at \url{https://gwosc.org/data/} and available either at the original sampling frequency of $16384$~Hz or downsampled to $4096$~Hz.
While there is not expected to be any signal content above a few thousand Hertz, for accurate reconstruction, it is important to start from the data sampled at $16382$~Hz since the details of the downsampling algorithm do impact the entire data set: using the pre-downsampled data can result in significant deviations between the reconstructed and original likelihoods.
For some events, the raw data is affected by transient non-Gaussian noise artefacts known as \emph{glitches}~\citep{Nuttall:2018xhi,Glanzer:2022avx,LIGO:2024kkz}.
In many cases, the glitches can be modelled and removed from the data using either a modelled Bayesian approach
\citep[\BAYESWAVE:][]{Pankow:2018qpo, Cornish:2020dwh, Chatziioannou:2021ezd, Hourihane:2022doe}, or a linear noise subtraction based on auxiliary witness channels \citep{Davis:2022ird}.
Therefore, to recreate the likelihood, the glitch-subtracted data should be used \citep[see, e.g.][for glitch-subtracted data for events from the first part of the fourth observing run]{ligo_scientific_collaboration_2025_16857060}.

\emph{Downsampling} -- For \ac{LVK} analyses, \BILBY uses the \soft{ResampleTimeSeries} downsampling routines implemented in \LALSUITE which implements a time-domain Butterworth filter \citep{butterworth1930theory}.
This filter was selected to match the downsampling performed by the tools used to generate the \ac{PSD} (discussed shortly).
The specific choice of sampling frequency is stored in the configuration file for the \BILBY analysis.

\emph{Windowing} --
The strain data used for inference is a subset of a longer time series of strain taken during the observation run.
Typically, a short section of data is cut out, containing the observable signal and sufficient padding on either side.
As a result, this time series is not periodic, which is required to approximately diagonalise the noise covariance matrix \citep[see the discussion in ][]{Talbot:2025vth}.
Therefore, a window factor is applied, gradually tapering the time series to zero at either end.
Following standard methods \citep{GWTC-4-methods}, a Tukey window \citep{harris2005use} is typically applied with a roll-off determined by a factor determining the fraction of the window that is tapered.
Before passing the data to likelihood, this window should be applied to the time-domain strain.

\emph{PSD} --
The GWTC analyses use \BAYESWAVE \citep{Cornish:2014kda,Littenberg:2014oda,Cornish:2020dwh,2024PhRvD.109f4040G} to estimate the on-source \ac{PSD} for each detector.
This is packaged within the data release and can be extracted and provided to the \BILBY likelihood.
Note that the data pre-processing should be identical for the analysis data and \ac{PSD} construction.

\emph{Waveform} --
The \BILBY likelihood uses a waveform approximant provided by \LALSIMULATION to simulate the frequency-domain signal $\mu(\theta)$ in \cref{eqn:likelihood}.
This is implemented within \BILBY in a \soft{WaveformGenerator} object that needs to be configured with the name of the waveform approximant to use, in addition to any configuration settings.

\emph{Explicit marginalization} --
The likelihood used for inference, at base, is the Whittle likelihood as given in \cref{eqn:likelihood}.
However, \BILBY implements several forms of explicit marginalisation, and these are routinely used to improve the convergence properties of the samplers \citep{GWTC-4-methods}.
As further described in \citet{Thrane:2018qnx, Romero-Shaw:2020owr}, after sampling \BILBY then reconstructs the full posterior distribution in post-processing.
Therefore, it is important to differentiate between the marginalised likelihood, which is used for sampling and the non-marginalised likelihood, which corresponds to the full posterior distribution.
In the data release, we show how both of these can be calculated.
In addition, we show that the log-likelihood that is stored in the data release corresponds to the marginalised likelihood.
The not-marginalised likelihood is not stored within the packaged data release, but can be calculated from the matched-filter and optimal \acp{SNR} that are stored (as shown in the data release).

\emph{Other configuration} --
While formally the likelihood for inference is a sum over the frequency bins of the Fourier transform, it is standard practise to omit data below a minimum frequency $f_{\rm min}$ and above a maximum frequency $f_{\rm max}$.
Typically, $f_{\rm min}$ is set to \qty{20}{\hertz}, except in cases where the data is affected by low-frequency non-Gaussian noise that cannot be subtracted, in which case a higher minimum frequency is chosen.
Meanwhile, the maximum frequency is by default set at a nominal choice above the maximum frequency of the signal.
In addition, \BILBY also has options to modify the choice of parameterisation.
For example, while the analyst may set a prior distribution on the geocentric arrival time \citep{GWTC-4-intro}, it is often advantageous for sampling to instead sample in the merger time relative to a fiducial detector.

\subsection{Prior distribution}
\BILBY provides an interface for defining the prior in terms of a \PYTHON dictionary.
The \emph{analysis} prior is packaged as part of the data release and can be easily reconstructed; this prior corresponds to the prior for the full posterior (e.g. after reconstruction of any marginalised parameters).
It is also helpful to define the \emph{run} prior that is used during stochastic sampling by \BILBY.
The run prior is created by modifying the analysis prior during the instantiation of the likelihood.
For example, if a parameter is explicitly marginalised, the run prior will replace the prior on this parameter with a delta-function prior at a fiducial value.
As with the likelihood, the log-prior stored in the result is from the run prior used during sampling.

In addition, some data releases apply a resampling of the posterior to utilise a cosmological prior on the luminosity distance.
Care must therefore be taken to ensure the prior is appropriately constructed to match the posterior samples.

\subsection{Calibration}
The strain has a known calibration uncertainty, which is represented in the frequency domain by a distribution on the phase and amplitude \citep[see, e.g.][]{Sun:2020wke}.
For standard \ac{LVK} analyses, \BILBY marginalises over the calibration uncertainty by taking as input a calibration envelope from which it constructs a spline approximation with an associated set of calibration parameters.
These are then sampled during inference to marginalise over the calibration uncertainty
\citep{Farr:2014aab, LIGOScientific:2016vlm}.
The calibration envelope is packaged as part of the data release and can be read in and used to construct a prior on the spline nodes.
A calibration model must also be provided to the interferometer instances that contain the strain data.
A demonstration of how this is done is provided in the data release.

\subsection{Demonstrations}
\label{sec:recon_demo}
\begin{figure*}
    \centering
    \includegraphics[width=0.45\linewidth]{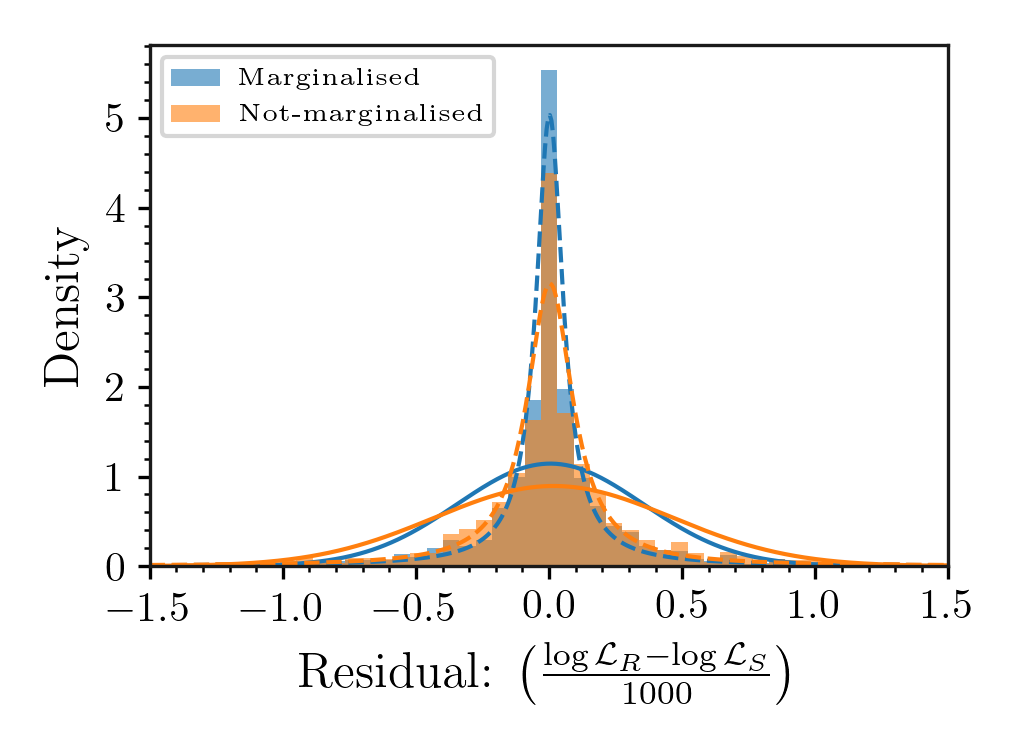}
    \includegraphics[width=0.45\linewidth]{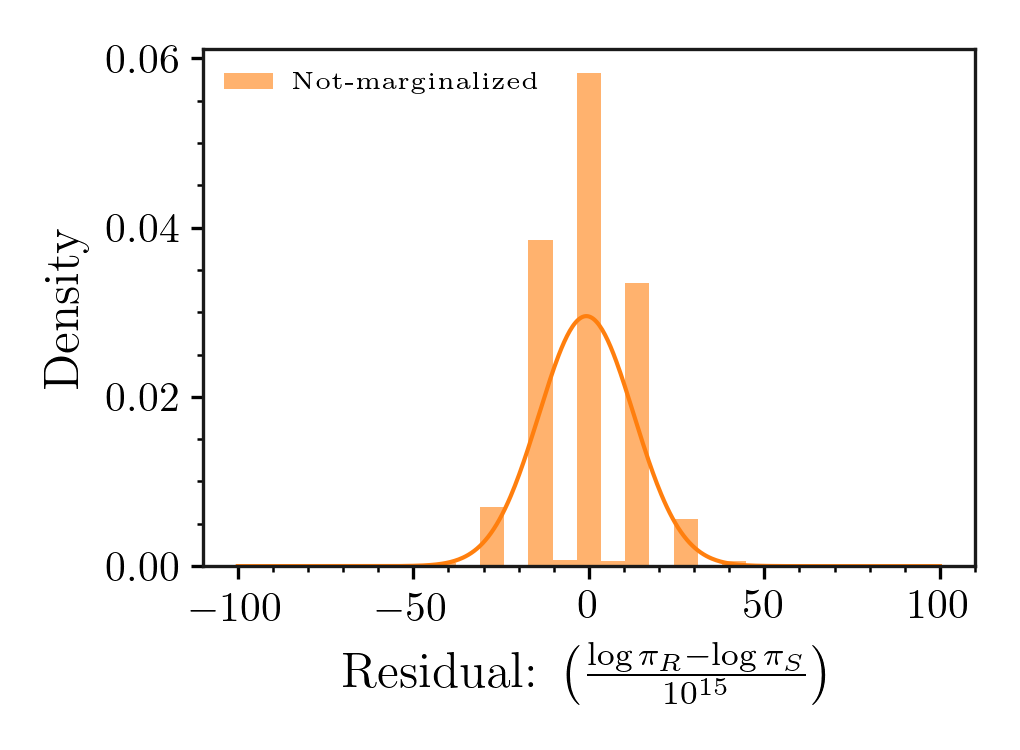} \\
        \includegraphics[width=0.45\linewidth]{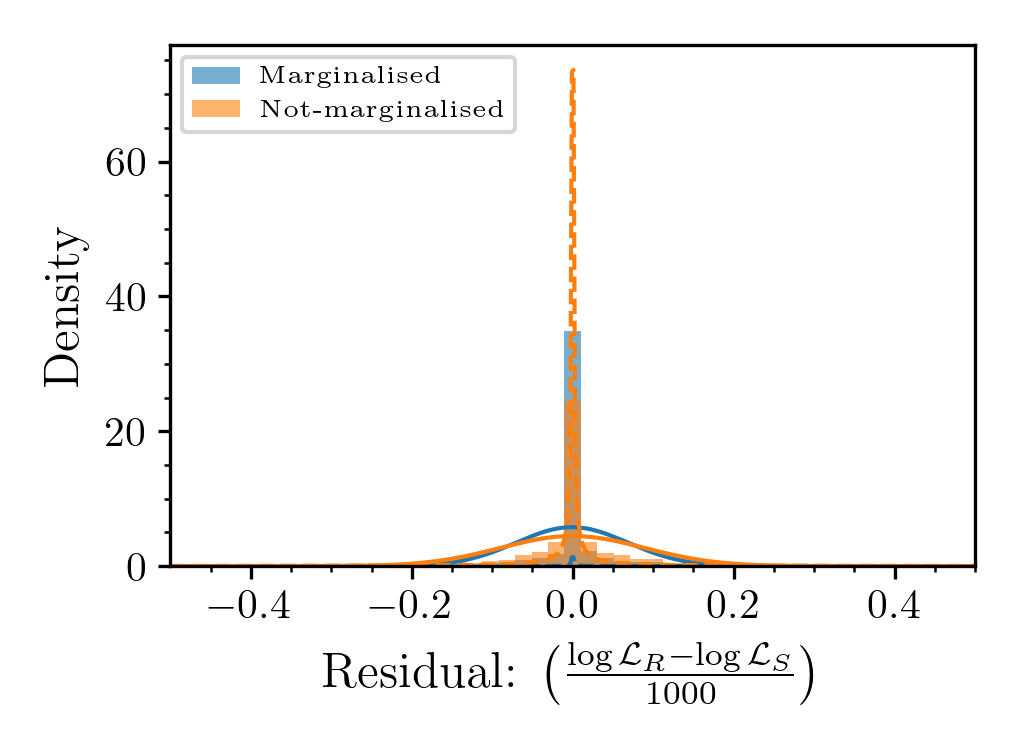}
    \includegraphics[width=0.45\linewidth]{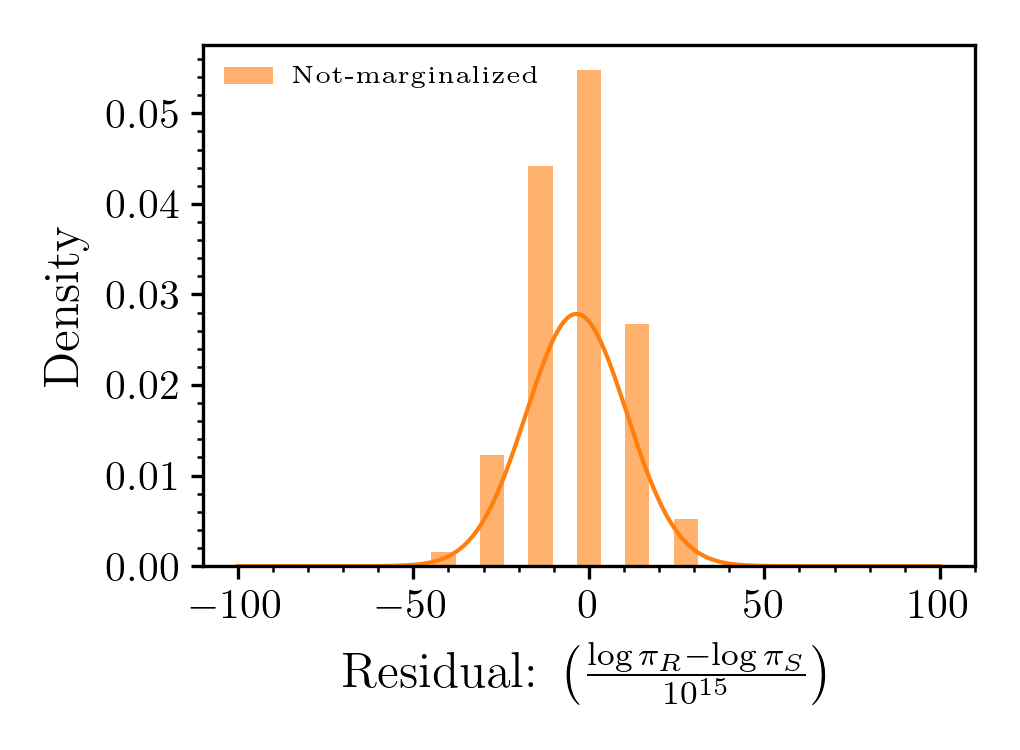} \\
    \caption{Histograms showing the residual difference between the reconstructed and stored log likelihood (left-hand column) and log-prior (right-hand column) for the GWTC-2.1 analysis of GW150914 (top row) and GWTC-4.0 analysis of GW230627\_015337 (bottom row).
    We fit a normal (solid curve) and Student's T distribution (dashed curve) to the likelihood residuals and present summaries of these fits in \cref{tab:fits}.
    }
    \label{fig:residuals}
\end{figure*}

In the data release, we provide notebooks demonstrating how to implement the setting described within this section for two analyses and calculate the reconstructed likelihood (both marginalised and non-marginalised) and prior.
We also show how to extract the stored values of the likelihood and prior.
We choose GW150914, the first observed binary black hole merger \citep{LIGOScientific:2016aoc}, and GW230627\_015337, a high-\ac{SNR} binary black hole observed during the first part of the fourth observing run \citep{GWTC-4-results}.
For GW150914, we use the GWTC-2.1 analysis \citep{LIGOScientific:2021usb, GWTC-2.1:PE}, which, amongst other things, utilised upgraded waveform models relative to the initial studies.
In \cref{fig:residuals}, we show histograms of the residual between the reconstructed and stored log-likelihood and log-prior;
in \cref{tab:fits}, we summarise fits to the residuals.
In both instances, the residuals are centred on zero, but while for the prior reconstruction we can achieve floating-point precision, for the likelihood, the residual values are scattered with random offsets below the level of one in a thousand.
For the log-likelihood, we fit both a normal distribution and a Student's T distribution, showing that the typical variations better fit the Student's T distribution with heavy tails.

While we do not extend this analysis to all observed events, we believe the notebooks in the data release should allow users to reconstruct the likelihood and prior for any observed event where \BILBY analyses have been produced by the \ac{LVK}.

\input{content/likelihood_prior_residuals_data}
\newcommand\sifmt[2]{{\sisetup{#1}\num{#2}}}
\newcommand\tro[1]{\sifmt{round-mode=figures,round-precision=1}{#1}}
\begin{table*}
    \centering
    \begin{tabular}{l|c|c|c|c}
         Event & Dist. & Range & Normal & Student's T \\\hline\hline
         \multirow{3}{*}{GW150914} & Marginalised & (\tro{\onefiveLMrangelow}, \tro{\onefiveLMrangehigh}) & (\tro{\onefiveLMnormMu}, \tro{\onefiveLMnormSigma}) & (\tro{\onefiveLMTMu}, \tro{\onefiveLMTNu}, \tro{\onefiveLMTSigma}) \\\
         & Not-marginalised & (\tro{\onefiveLNrangelow}, \tro{\onefiveLNrangehigh}) & (\tro{\onefiveLNnormMu}, \tro{\onefiveLNnormSigma}) & (\tro{\onefiveLNTMu}, \tro{\onefiveLNTNu}, \tro{\onefiveLNTSigma}) \\\
         & Prior & (\tro{\onefivePrangelow}, \tro{\onefivePrangehigh}) & (\tro{\onefivePnormMu}, \tro{\onefivePnormSigma}) & --- \\ \hline
         \multirow{3}{*}{GW230627\_015337} & Marginalised & (\tro{\twothreeLMrangelow}, \tro{\twothreeLMrangehigh}) & (\tro{\twothreeLMnormMu}, \tro{\twothreeLMnormSigma}) & (\tro{\twothreeLMTMu}, \tro{\twothreeLMTNu}, \tro{\twothreeLMTSigma}) \\\
         & Not-marginalised & (\tro{\twothreeLNrangelow}, \tro{\twothreeLNrangehigh}) & (\tro{\twothreeLNnormMu}, \tro{\twothreeLNnormSigma}) & (\tro{\twothreeLNTMu}, \tro{\twothreeLNTNu}, \tro{\twothreeLNTSigma}) \\\
         & Prior & (\tro{\twothreePrangelow}, \tro{\twothreePrangehigh}) & (\tro{\twothreePnormMu}, \tro{\twothreePnormSigma}) & --- \\\
    \end{tabular}
    \caption{Summary statistics for the distribution of the residuals presented in \cref{fig:residuals}.
    We give the minimum and maximum of the range, the fitted mean and standard deviation of a normal distribution, and the fitted mean, standard deviation, and degree of freedom for the Student's T distribution (but this is not applied to the prior residual).
    For the likelihood residuals, all values are scaled by $1000$, while for the prior residuals, values are scaled by $10^{15}$.
    }
    \label{tab:fits}
\end{table*}

%% file: content/likelihood_prior_residuals_data.tex
\newcommand{\onefiveLMrangelow}{-3.8209945952303315}
\newcommand{\onefiveLMrangehigh}{4.046571464641602}
\newcommand{\onefiveLMnormMu}{0.0048082148828143545}
\newcommand{\onefiveLMnormSigma}{0.34877312719850956}
\newcommand{\onefiveLMTMu}{0.0020471429641639675}
\newcommand{\onefiveLMTNu}{0.9451576037283338}
\newcommand{\onefiveLMTSigma}{0.0625328334925736}
\newcommand{\onefiveLNrangelow}{-3.8054937232345765}
\newcommand{\onefiveLNrangehigh}{5.002172084459744}
\newcommand{\onefiveLNnormMu}{0.017448105721257434}
\newcommand{\onefiveLNnormSigma}{0.44612123786092867}
\newcommand{\onefiveLNTMu}{0.003157418091364096}
\newcommand{\onefiveLNTNu}{1.090441716182796}
\newcommand{\onefiveLNTSigma}{0.10260152988301352}

\newcommand{\onefivePrangelow}{-42.63256414560601}
\newcommand{\onefivePrangehigh}{42.63256414560601}
\newcommand{\onefivePnormMu}{-0.7105427357601002}
\newcommand{\onefivePnormSigma}{13.505920362421252}

\newcommand{\twothreeLMrangelow}{-0.555142630616956}
\newcommand{\twothreeLMrangehigh}{0.44784223831584313}
\newcommand{\twothreeLMnormMu}{-0.0008242876326960413}
\newcommand{\twothreeLMnormSigma}{0.06903938107973431}
\newcommand{\twothreeLMTMu}{1.5093230705763324e-12}
\newcommand{\twothreeLMTNu}{0.26230390129468606}
\newcommand{\twothreeLMTSigma}{9.266943721383815e-11}
\newcommand{\twothreeLNrangelow}{-0.5415481583668225}
\newcommand{\twothreeLNrangehigh}{0.4436625426365026}
\newcommand{\twothreeLNnormMu}{-0.0003652334859509665}
\newcommand{\twothreeLNnormSigma}{0.08930183707539167}
\newcommand{\twothreeLNTMu}{1.9863714753297437e-06}
\newcommand{\twothreeLNTNu}{0.25900542461990017}
\newcommand{\twothreeLNTSigma}{0.0004627512644043015}

\newcommand{\twothreePrangelow}{-42.63256414560601}
\newcommand{\twothreePrangehigh}{42.63256414560601}
\newcommand{\twothreePnormMu}{-3.510081114654895}
\newcommand{\twothreePnormSigma}{14.324026964120064}

%% file: content/3-resampling.tex
\section{Resampling the posterior}
\label{sec:resampling}

In this Section, we will introduce the formalism of resampling in the context of applications to a set of released posterior samples $\{\theta_i\}$.
We note that this presentation is intended to be specific to the use-case of gravitational-wave astronomy; for a more general guide, see, e.g. \citet{mackay2003information}.

Given a sample $\theta_i$, we can calculate the unnormalised posterior density 
\begin{align}
   p(\theta_i) = \likelihood(\data | \parameters_i, \model) \prior(\parameters_i | \model) \,,
   \label{eqn:posterior_density}
\end{align}
where we use the term density to distinguish it from the distribution indicated in \cref{eqn:posterior} and drop the conditional statements to simplify the notation.
In the language of resampling, we refer to $p(\theta)$ as the \emph{proposal} distribution.
It is assumed that the set of samples $\{\theta_i\}$ is properly drawn from the distribution; if this is not the case, for example, if the stochastic sampler has not converged, then resampling will inherit the bias.

Next, we wish to resample to a \emph{target} distribution $p'(\theta)$ with an associated density:
\begin{align}
   p'(\theta_i) = \likelihood'(\data' | \parameters_i, \model') \prior'(\parameters_i | \model') 
\end{align}
where the prime could mean a variation in the choice of likelihood $\likelihood \rightarrow \likelihood'$ (which could include, for example, changes to the data processing or assumptions about the noise), a change in the generative model $\model \rightarrow \model'$, or a change in the choice of prior $\prior \rightarrow \prior'$ (under some interpretations, the prior is part of the model, but here we distinguish it as a separate choice to make it clear if instead the generative model alone is changing).
Note that while it is possible to change more than one of these at a time, it would be advisable to change one at a time, at least initially, to understand their individual impact.

Given the proposal distribution and target distribution, we define a generalised weight function
\begin{equation}
    w(\theta) = \frac{p'(\theta)}{p(\theta)}\,.
    \label{eqn:generalised_weights_function}
\end{equation}
Then, given the set of samples and the computed proposal and target values, we construct a set of weights calculated as the ratio of the densities:
\begin{align}
    w_i = w(\theta_i) = \frac{p'(\theta_i)}{p(\theta_i)}\,.
    \label{eqn:weights}
\end{align}
The goal of the following subsections is to outline different strategies to resample the posterior using these weights as an alternative to reanalysis, applying stochastic sampling directly to estimate $p'(\theta | \model, \data)$.
We note that in some applications, the calculation of the weights greatly simplifies.
For example, if only the prior changes, then the likelihood terms cancel.
However, our examples are given with the fully general case to enable the reader to apply it in arbitrary contexts.

\subsection{Rejection sampling}
\label{sec:rs}
\begin{figure}
    \centering
    \includegraphics[]{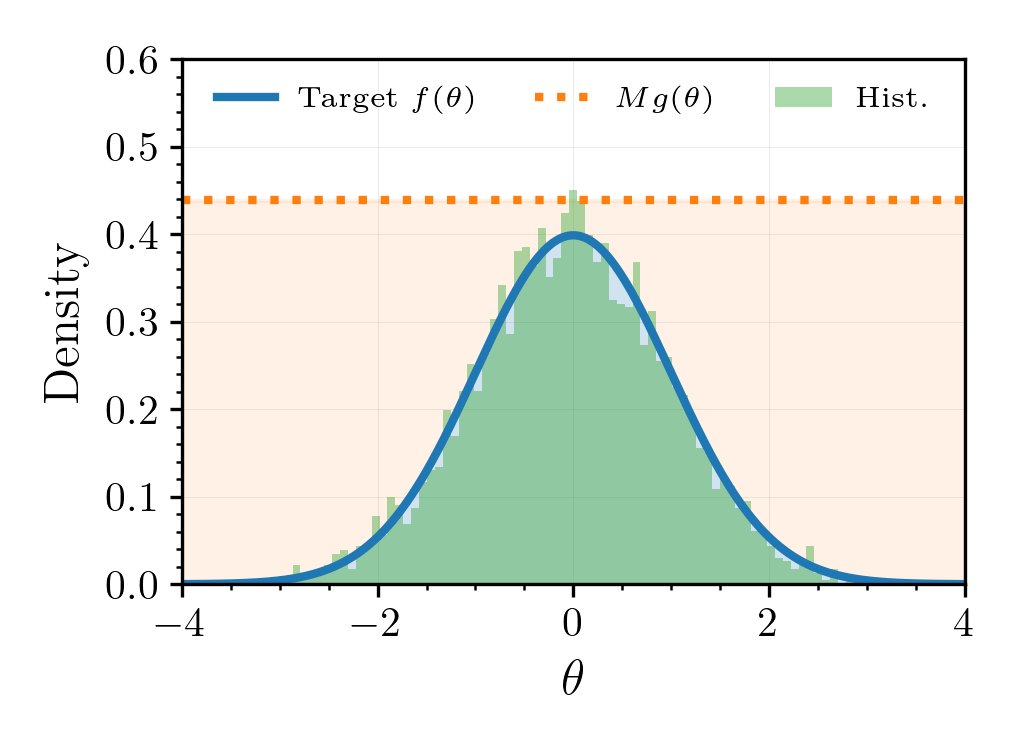}
    \caption{A simple illustration of rejection sampling for a standard normal target density $f(x)$ and uniform proposal density $g(x)$.
    We plot the target density (blue solid curve), scaled proposal density (orange dashed curve), and a histogram of the accepted samples from \num{10000} iterations of the algorithm.
    Proposal samples, $\parameters_i$ drawn from $p(\parameters)$ are accepted with a probability proportional to $p'(\parameters)/Mp(\parameters)$.
    Visually, in the figure above, this is the ratio of the solid blue curve to the orange dashed line.
    }
    \label{fig:rejection_sampling}
\end{figure}

\Ac{RS} is a classical Monte Carlo technique which draws samples from a proposal distribution $p'(\parameters)$ using a generating distribution $p(\parameters)$ and a constant $M$ which bounds the ratio:
\begin{align}
    M \ge \mathrm{sup}_\parameters \frac{p'(\parameters)}{p(\parameters)}\,,
    \label{eqn:M}
\end{align}
where $\mathrm{sup}_\parameters f(\parameters)$ is the \emph{supremum} of $f(\parameters)$ for all possible values of $\parameters$; in the context of continuous functions with a maximum, the supremum is the maximum.
\ac{RS} starts by drawing a sample $\parameters_i \sim p(\parameters)$ and then defining an acceptance probability for the sample $p'(\parameters_i)/M p(\parameters_i)$ which is proportional to how well the proposed sample matches the target distribution $p'(\parameters)$ normalised by the envelope $Mp'(\parameters)$.
Algorithmically, this is implemented by first drawing $u \sim \mathrm{Unif}(0, 1)$ and then accepting the sample if
\begin{align}
    u < \frac{p'(\theta_i)}{M p(\theta_i)} \,,
\end{align}
and rejecting the sample otherwise.
After applying this test to all samples in $\{\theta_i\}$, we end up with a resampled posterior distribution $\{\theta_i'\}$ under the assumptions of the primed distribution.
To illustrate the basic idea, in \cref{fig:rejection_sampling} we provide a plot from a simple example applying rejection sampling for a uniform proposal distribution and standard normal target distribution.

The bound $M$ ensures that $p'(\parameters) \le M p(\parameters)$ for all possible values of $\parameters$ and is ideally chosen to be as small as possible while respecting \cref{eqn:M}.
Visually, in \cref{fig:rejection_sampling}, the closer the $M g(x)$ curve is to the peak of $f(\parameters)$, the more efficient the sampler, as the smaller the rejection region is.
More formally, if we define \ac{RS} efficiency as
\begin{align}
    \eta_{\rm RS} = 
    \frac{N'}{N}\,,
    \label{eqn:eta_RS_def}
\end{align}
where $N'$ is the number of accepted samples.
Then the efficiency can be calculated from averaging the acceptance ratio:
\begin{align}
    \eta_{\rm RS} = 
    \expect{\frac{p'(\parameters)}{M p(\parameters)}}_{p(\parameters)}
    =\frac{1}{M} \int_\parameters p'(\parameters) \,d\parameters\,,
    \label{eqn:eta_RS_calc}
\end{align}
where we introduce
$\expect{g(x)}_{f(x)}$ as the expectation of $x$ under $f(x)$:
\begin{align}
    \expect{g(x)}_{f(x)} = \int g(x) f(x)\, dx\,.
\end{align}
Hence, the smallest value of $M$ that satisfies \cref{eqn:M} maximises the efficiency.
In practice, the optimal value of $M$ is unknown in advance and must be chosen instead.
Typically, this is done by selecting
\begin{align}
    M = \mathrm{max}(w_i)
    \label{eqn:Mmax}
\end{align}
which ensures the condition is met over the set of samples.
However, it does not ensure the condition is met in general.
(From \cref{fig:rejection_sampling}, it can be seen that if \cref{eqn:M} is not met, the posterior will be clipped for all \parameters where $p'(\parameters) > M p(\parameters)$).

If $M$ is approximated by \cref{eqn:Mmax}, the acceptance criteria is
\begin{align}
    u < \frac{w_i}{\mathrm{max}(w_i)}\,,
\end{align}
which is the standard implementation often used in the field.

With this choice of $M$, the efficiency can be computed directly from the weights as
\begin{align}
    \eta_{\rm RS} = \frac{\expect{\{w_i\}}}{\mathrm{max}(\{w_i\})}\,,
\end{align}
where $\expect{\{w_i\}}$ is the mean of the weights.
This demonstrates that efficient rejection sampling requires the distribution of weights to be compact, with a maximum not much larger than the mean.
If, on the other hand, the maximum weight is much larger than the mean, sampling will be inefficient.

We can further explore the efficiency by considering the case when the weights are log-normally distributed (i.e. that the distribution of $\ln w_i$ is approximately normal) with a mean of $\mu$ and variance $\sigma^2$.
Then, the mean of the weights is given by $\mathrm{exp\left(\mu + \sigma^2 / 2\right)}$ while we can approximate the maximum of a set $N$ weights by $\max(\{\parameters_i\})\approx \mu + \sigma \sqrt{2\mathrm{log}N}$ yielding
\begin{align}
    \eta_\mathrm{RS} \approx \mathrm{exp}\left( -\sigma\sqrt{2\ln N}+ \frac{\sigma^2}{2}\right)\,.
    \label{eqn:eta_RS_A}
\end{align}
This approximation breaks down for $\sigma \gtrsim 1$, but provides a useful estimate of the expected efficiency for $\sigma < 1$.
Furthermore, for $\sigma \ll 1$ we obtain
\begin{align}
    \eta_\mathrm{RS} \approx 1 - \sigma \sqrt{2\ln N}\,.
    \label{eqn:eta_RS_B}
\end{align}

We can use this approximation to provide a bound on the required accuracy of likelihood reconstruction in \cref{sec:reconstructing}, assuming they are log-normally distributed (in practise we find a Student's T distribution is a better fit to the residual of the log-likelihood weights, but it nevertheless provides an approximate bound).
To ensure the sampling efficiency is impacted at the 1\% level or less (for $N=10000$ samples), we require $\sigma < 0.0035$.
Comparing this to the distributions in \cref{fig:residuals}, we can conclude that the reconstructions are sufficiently accurate to have a negligible impact on the resampling efficiency.
Of course, in practise the distribution will depend on the changed assumptions and will likely not be log-normally distributed.
Nevertheless, the approximations can provide a useful heuristic to predict the performance. 

In the straightforward approach described above, the number of samples produced by \ac{RS}, $N'$, is determined by the acceptance ratio (see \cref{eqn:eta_RS_def}).
However, we note that it is also possible to repeatedly apply the \ac{RS} algorithm until a pre-defined number of samples is produced, $N_t$.
If $N_t > N'$, the resulting sample set will no longer be \ac{IID}, but this technique is nevertheless useful for smoothing histograms (however, see the discussion in the next section for alternatives).

\subsection{Importance sampling}
\label{sec:is}
\Ac{IS} is an alternative Monte Carlo approach which, in its textbook definition, rather than creating a new set of equally-weighted samples (as done by \ac{RS}), instead utilises the samples and associated weights directly.
For example, to visualise the re-weighted posterior distribution, a histogram could be produced using the weights to modify each sample's contribution to the histogram density.
Or alternatively, for summary statistics, the weights can be used to modify the contribution from each sample.
E.g., the weighted mean of a distribution can be computed as
\begin{align}
    \expect{\parameters}_{p(\parameters)} = \int p(\parameters) \parameters\,d\parameters \approx \sum_{i} \bar{w}_i \theta_i \,,
\end{align}
where $\bar{w}_i$ are the normalised weights
\begin{align}
    \bar{w}_i = \frac{w_i}{\sum_i w_i}\,.
    \label{eqn:normalised_weights}
\end{align}

There is no direct way to compare the efficiency of \ac{RS} and \ac{IS} because they are fundamentally different approaches.
However, it can nevertheless be useful to consider a different measure of efficiency for \ac{IS}, the Kish \ac{ESS}:
\begin{align}
    \Ness = N \frac{\expect{w}^2}{\expect{w^2}} = \frac{\left(\sum_i w_i\right)^2}{\sum_i{w_i}^2}
    =\frac{1}{\sum_i \bar{w}_i^2}\,,
    \label{eqn:ess}
\end{align}
where we give three common and equivalent definitions: in terms of the mean of the weights and squared weights, in terms of the weights, and in terms of the normalised weights.

The \ac{ESS} was introduced \citep{kish1965survey} as a measure of the loss of precision due to unequal weights. Specifically, it can be interpreted as the number of samples that would need to be drawn from the target distribution to match the variance of the estimate taken from the weighted distribution.
However, the definitions given in \cref{eqn:ess} are an approximation that can be computed from the samples themselves (see \citet{kong1992note} for the derivation and \citet{elvira2022rethinking} for a detailed discussion).

We then define the \ac{IS} efficiency as
\begin{align}
    \eta_{\rm IS} = \frac{\Ness}{N}\,,
    \label{eqn:eta_IS_def}
\end{align}
and hence, considering \cref{eqn:ess} and assuming the weights are log-normally distributed, the efficiency of \ac{IS} can then be estimated from
\begin{align}
    \eta_\textrm{IS} = \frac{\expect{w}^2}{\expect{w^2}} = \exp{\left(-\sigma^2\right)}\,.
    \label{eqn:eta_IS_approx}
\end{align}

\begin{figure}
    \centering
    \includegraphics[width=\linewidth]{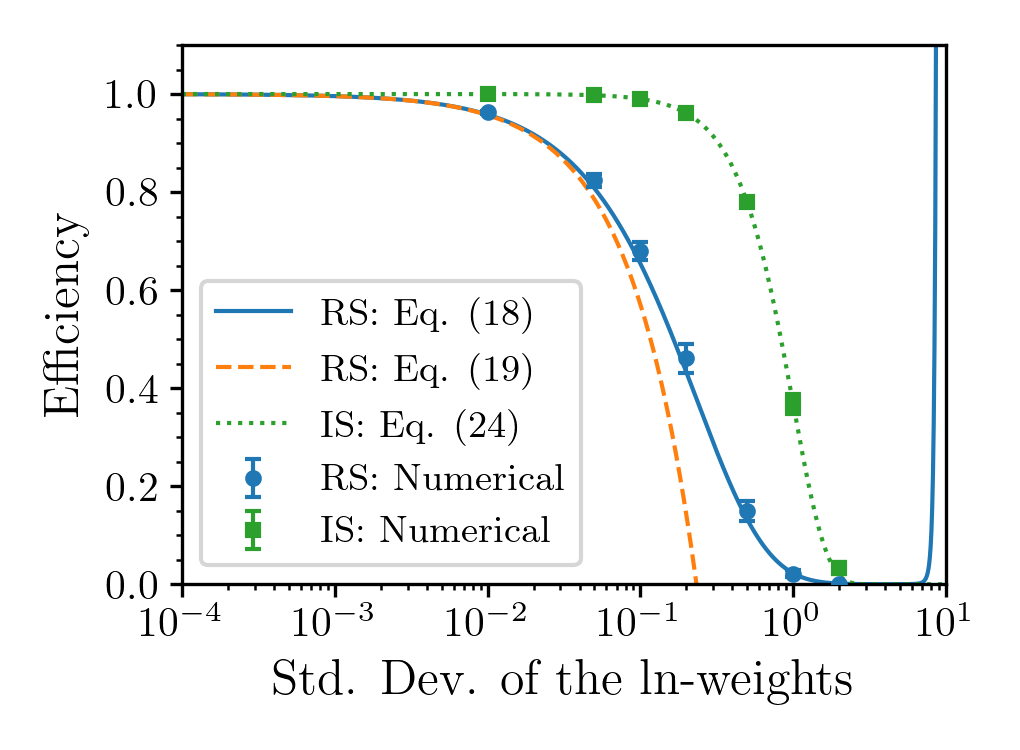}
    \caption{The efficiency of the \ac{RS} and \ac{IS} approaches as predicted by \cref{eqn:eta_RS_A}, \cref{eqn:eta_RS_B} and \cref{eqn:eta_IS_approx}.
    To numerically validate the theoretical predictions, we add numerical results from a toy model where the target distribution is a standard normal $x\sim N(0, 1)$, and the proposal is a shifted normal distribution $x'\sim N(\mu, 1)$.
    For this setup, the log-weight is $\mu x - \mu^2/2$, which has standard deviation $\sigma = |\mu|$ when sampling $x \sim N(0,1)$.
    By varying $\mu$, we can control $\sigma$ and compare empirical efficiency measurements against the theoretical predictions, demonstrating agreement.
    For further details, please see the software behind the figure in the data release \citep{data_release}.
    }
    \label{fig:efficiency}
\end{figure}

In \cref{fig:efficiency}, we compare the predicted efficiency of the \ac{RS} and \ac{IS} approaches assuming the weights are log-normally distributed.
We also add a comparison with a numerical example, validating the predictions.
This suggests that \ac{IS} is more efficient for moderate values of $\sigma$, though both of course become highly inefficient as $\sigma$ approaches unity.
However, we note that these efficiencies should not be taken at face value, as they present fundamentally different statistical concepts: while $\eta_{\rm RS}$ is the acceptance rate of the \ac{RS} algorithm, $\eta_{\rm IS}$ measures the ratio of the effective sample size of the weighted sample set.
Nevertheless, they help us to identify a key difference between \ac{RS} and \ac{IS}, namely a trade-off between smoothness and noise:
while \ac{RS} produces pure equally-weighted samples, the level of noise (given by the square root of the number of accepted samples) is larger compared to \ac{IS}, which keeps the entire set of samples (maximising the smoothness of the resulting histograms).

A common issue with \ac{IS} is the requirement to package the weights alongside the samples.
Since many downstream methods assume a set of equally weighted samples, this can be problematic.
An alternative is therefore to instead apply multinomial-\ac{IS} in which we draw $N_t$ samples (with replacement) from the original samples $\{\theta_i\}$ with probabilities given by the normalised weights in \cref{eqn:normalised_weights}.
$N_t$ could either be taken to be the size of the original sample set, or a more conservative choice is to take $\Ness$.
In either case, the resulting resampled posterior $\{\parameters_i'\}$ is no longer \ac{IID} (because it will contain repeated samples), but it can present smoother histograms compared to \ac{RS}.
We note that multinomial-\ac{IS} is identical in behaviour to repeated application of \ac{RS} (as discussed at the end of the last Section).
However, multinomial-\ac{IS} is computationally more efficient.

\subsection{Pareto-smoothing}
\label{sec:psis}

When the distribution of weights has a heavy right tail, both \ac{RS} and \ac{IS} can be highly inefficient.
This typically occurs because the proposal and target distribution differ.
A technique commonly applied in the field of statistics (but not yet common in gravitational-wave astronomy, thourgh see \citet{Mould:2025dts}) is the use of Pareto-smoothing, which modifies the right tail of the weight distribution to stabilise resampling and reduce the error of \ac{IS} estimates \citep{vehtari2024pareto}.
The approach was developed as an extension of \ac{IS} and named \ac{PSIS}.
A generalised Pareto distribution is fitted to the logarithm of the largest weights (typically assigned by the user, with a typical value of the largest 20\%).
These weights are then replaced with the expected order statistics of the fitted distribution, and then the entire set of weights is re-normalised.
This process smoothes the tails of the distribution, and enables the entire set of samples to be used in \ac{IS} with improved efficiency.
Pareto smoothing does induce a small bias, but this is often acceptable since only the tail of the distribution is affected.
However, the algorithm yields a diagnostic $\hat{k}$, the tail index of the fitted distribution, that can be used to assess the impact.
For $\hat{k}<0.5$, the bias is expected to be negligible, and results can be trusted.
For $0.5 \le \hat{k} < 0.7$, care should be taken.
And for $\hat{k} > 0.7$, the estimates are unstable or have a high degree of bias.
$\hat{k}$ can also be negative, indicating that the tail of the weight distribution decays faster than a Pareto distribution would, and smoothing will have a negligible impact.

\begin{figure*}
    \centering
    \includegraphics[width=\textwidth]{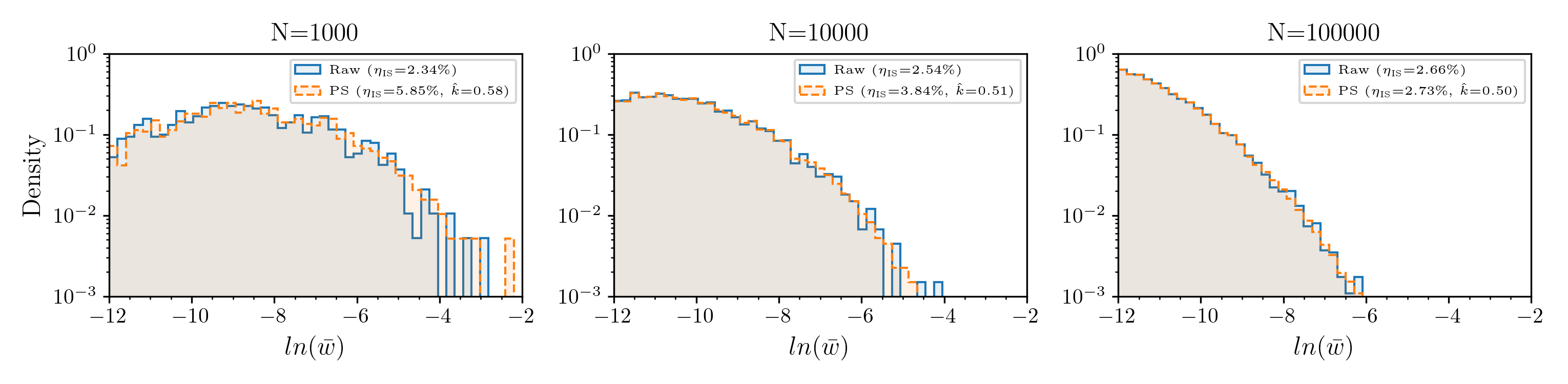}
    \caption{Histograms of simulated weights (Raw: blue solid curve) and Pareto-smoothed versions (PS: orange dashed curves).
    The weights are simulated from a distribution $w \sim e^{n}$, where $n$ is a random variable drawn from a normal distribution with zero mean and standard deviation of 2.
    We plot histograms of the natural logarithm of the normalised weights and truncate the histogram at a lower edge to focus on the right-hand tail.
    In the legend, we provide efficiency metrics and the Pareto-smoothing diagnostic $\hat{k}$. 
    }
    \label{fig:pareto}
\end{figure*}

In \cref{fig:pareto}, we present a demonstration of Pareto smoothing for some simulated weights with a heavy right tail using the \soft{arviz} implementation \citep{arviz_2019}.
We vary $N$, the number of samples, to show that smoothing is most effective when $N$ is small, such that there is significant variance in the tail.

While \ac{PSIS} is predominantly used in the context of \ac{IS}, it can also be used to smooth the weights for \ac{RS}.
We refer to this as \ac{PSRS}.
We note that, like \ac{PSIS}, the samples produced from \ac{PSRS} will be biased as the distribution is an approximation to the target distribution.
However, often this bias is worthwhile for the improvement in efficiency.

\subsection{Evidence estimation}
\label{sec:evidence}
Typically, alongside the initial sample set $\{\parameters_i\}$, the Bayesian evidence, \cref{eqn:evidence}, is also estimated during the stochastic sampling process.
When resampling, we can use the weights, \cref{eqn:weights}, to obtain a Bayes factor between the primed and unprimed evidence values as follows.
First, we write the definition of the primed evidence
\begin{align}
    \evidence'(d' | \model') = \int \likelihood'(d' | \parameters, \model') \prior'(\parameters)\, d\theta\,.
\end{align}
Then multiply the integrand by the proposal likelihood and prior
\begin{align}
    \evidence'(d' | \model') &= \int \frac{\likelihood'(d' | \parameters, \model') \prior'(\parameters | \model)}{\likelihood(d | \parameters, \model) \prior(\parameters | \model)} \likelihood(d | \parameters, \model) \prior(\parameters | \model)\,d\theta\,. \\
    &=
    \int w(\theta) \likelihood(d | \parameters, \model) \prior(\parameters | \model)\,d\theta\,.
\end{align}
where we have used the definition of the generalised weight function, \cref{eqn:generalised_weights_function}.
Now, we recognise that $\likelihood(d | \parameters, \model) \prior(\parameters | \model)=\evidence(\data | \model) p(\parameters | \data, \model)$ and therefore
\begin{align}
    \evidence'(d' | \model') &= \evidence(\data | \model) \int w(\parameters) p(\parameters | \data, \model)\, d\theta\,,
\end{align}
and so finally, if we have $N$ samples $\{\parameters_i\}$ drawn from $p(\theta | d, M)$, then we can replace the integral with a Monte Carlo approximation and hence form a Bayes factor
\begin{align}
    \BF = \frac{\evidence'(d' | \model')}{\evidence(\data | \model)} &= \frac{1}{N} \sum_i^N w_i = \expect{w} \,.
    \label{eqn:BF}
\end{align}
(Note: a similar derivation can be found in \citet{Payne:2019wmy} when only the likelihood is changing).
For numerical stability, it is usually better to estimate the logarithm of the Bayes factor from
\begin{align}
    \ln\BF = \mathrm{LSE}\left(\{\ln w_i\}\right) - \ln(N)\,,
    \label{eqn:logBF}
\end{align}
where $\mathrm{LSE}(\{x_i\}) \equiv \ln(\sum_i \exp(x_i))$ is the logarithm of the sum of the exponentials, a function for which computational libraries such as \SCIPY \citep{2020SciPy-NMeth} provide convenient methods that preserve numerical stability.

Given a set of weights, \cref{eqn:BF} can be used to perform a comparison of the evidence under the two differing assumptions.
If an estimate of $\evidence(\data | \model)$ is available, then multiplying by the Monte Carlo average provides a new estimate of $\evidence'(d' | \model')$.

To estimate the uncertainty in the Bayes factor introduced by the Monte Carlo sum, we first note that the standard error on the Bayes factor is given by
\begin{align}
    \sigma_\avew = \frac{\sigma_w}{\sqrt{N}}
    \label{eqn:sigma_avew}
\end{align}
where $\sigma_w^2$ is the true variance of the weights.

For a given set of weights, computing $\sigma_w$ and inserting this into \cref{eqn:sigma_avew} can be used to estimate the uncertainty on the Bayes factor calculated in \cref{eqn:BF}.
If instead, the log-Bayes factor is calculated, \cref{eqn:logBF}, then propagating the uncertainty:
\begin{align}
    \sigma_{\ln\BF} = \frac{\sigma_w}{\BF\sqrt{N}}\,.
\end{align}

Starting from \cref{eqn:sigma_avew}, we can understand the performance in more detail by estimating $\sigma_w$ (the true variance of the weights) with the sample variance:
\begin{align}
    \sigma_w^2 & \approx \frac{1}{(N-1)} \sum_i \left(w_i - \langle w \rangle\right)^2 \\
    & = \frac{1}{N-1}\left(\sum_i w_i^2 - N \avew^2\right)\,,
    \label{eqn:expA}
\end{align}
where we have rearranged the expression in the second line.
Next, recalling the definition of the \ac{ESS}, we can rewrite \cref{eqn:ess} as
\begin{align}
   \Ness = \frac{N^2 \avew^2}{\sum_i {w_i}^2}\,.
\end{align}
Now combining with \cref{eqn:expA}, we can identify that
\begin{align}
    \sigma_w^2 = \avew^2 \left(\frac{N}{\Ness} - 1\right)\,
\end{align}
and hence the relative error in the Bayes factor, \cref{eqn:BF}, is
\begin{align}
    \frac{\sigma_\avew}{\avew} = \frac{1}{\sqrt{\Ness}} \sqrt{1 - \frac{\Ness}{N}}
    = \frac{1}{\sqrt{\Ness}} \sqrt{1 - \eta_{\rm IS}}\,.
\end{align}
This expression shows that when the efficiency is high, the variance in the estimator of the Bayes factor is small, and we recover the usual variance scaling $\propto 1 / \sqrt{N}$.
Meanwhile, when the efficiency is small, such that $\Ness \ll N$, the variance instead scales as $\propto 1/\sqrt{\Ness}$.
In other words, not only is the \Ness useful in determining the efficiency of \ac{IS}, it also determines the accuracy of the Monte Carlo integration approach to estimating the Bayes factor, \cref{eqn:BF}.

%% file: content/4-examples.tex
\section{Examples}
\label{sec:example}

In this Section, we will provide several simple examples of the use of resampling.
We begin with a simple toy model that investigates the sampling approaches discussed in \cref{sec:resampling}, then move on to demonstrations from gravitational-wave astronomy.
Details of how to reproduce these examples can be found in the data release \citep{data_release}.

\subsection{Validating and exploring resampling methods}
\label{sec:toymodel}
We now demonstrate the application of the resampling methods described in \cref{sec:resampling} to a toy model.
Of course, the methods are already well-validated within the statistics literature, but our goal is to use tools familiar to gravitational-wave astronomers to demonstrate the qualities of the methods.

To construct a toy problem, we consider a one-dimensional inference problem in which a variable $x$ is measured to have a posterior distribution $p(x)$ that follows a truncated normal distribution bounded on the unit interval.
We simulate an ensemble of measurements of $x$, drawing true values $\mu_x$ from a population prior distribution $\mu_x=\mathrm{Unif}(0, 1)$.
To each true value, we then simulate measurement error by adding a random variable $\Delta\mu_x\sim \mathrm{Norm}(\delta\mu_e, \sigma_e)$, where $\delta\mu_e$ and $\sigma_e$ are the bias and standard deviation in the measurement error.
We then draw samples from the simulated posterior distribution, $x\sim \mathrm{Norm}(\mu_x + \delta\mu_e, \sigma_p)$ where we introduce $\sigma_p$ as the standard deviation of the Gaussian posterior.
For unbiased posteriors, the mean of the measurement error is zero $\delta\mu_e=0$ and the standard deviations of the measurement error $\sigma_e$ and simulated posterior distributions $\sigma_p$ are identical, $\sigma_e=\sigma_p$.
However, if $\delta\mu_e \neq 0$, this simulates bias.
Meanwhile, if $\sigma_p \neq \sigma_e$, this simulates a scale-bias (over-constrained or under-constrained).

\begin{figure}
    \centering
    \includegraphics[width=\linewidth]{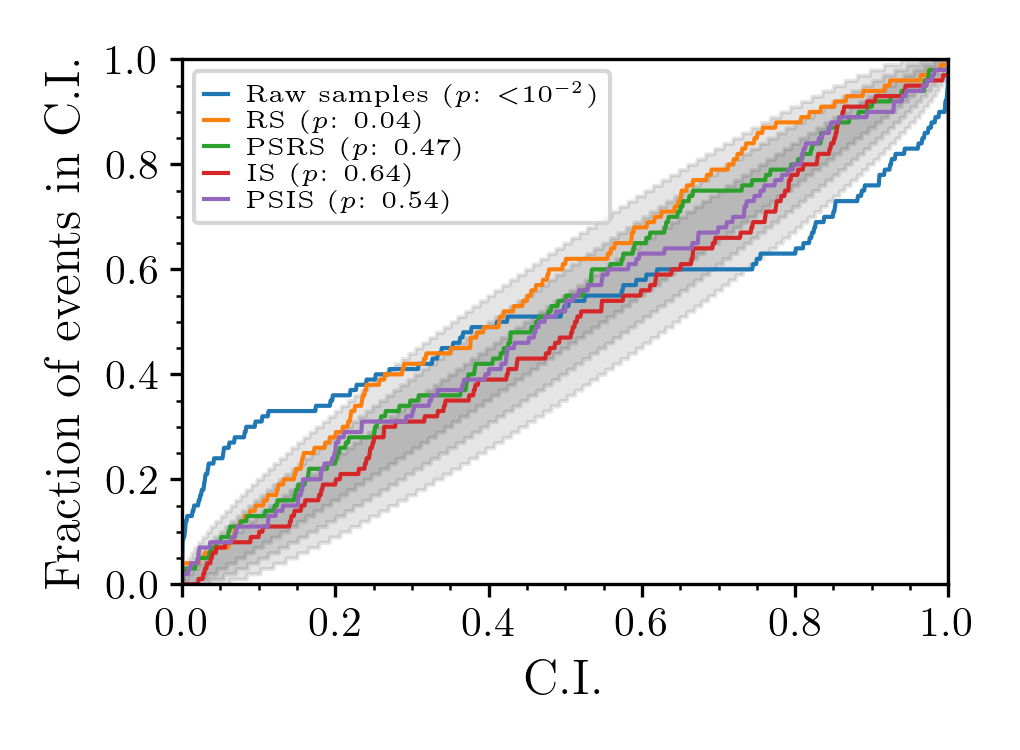}
    \caption{A \ac{PP} test for the concocted under-constrained posterior distributions with $\sigma_e/\sigma_p=2$ as described in \cref{sec:toymodel}.
    In the legend, we give the $p$-value of the \ac{PP} test applied to the raw samples (which fail at a threshold of $0.01$) while all four resampled posterior sample sets pass at this threshold.
    In \cref{sec:pp}, we describe the details of how the \ac{PP} test is applied and how the $p$-values provided in the legend are calculated.
    }
    \label{fig:pp}
\end{figure}

We concoct a scenario in which the posteriors are under-constrained $\sigma_e / \sigma_p=2$, but unbiased, with $\delta\mu_e=0$, and then generate many realisations of the posterior distribution.
In \cref{fig:pp}, we calculate a \ac{PP} test \citep{Sidery:2013zua, Veitch:2014wba, Romero-Shaw:2020owr} showing that the concocted posterior distributions are scale-biased (do not follow the expected 1-to-1 relation) and fail a $p$-value test.
We discuss the specifics of the \ac{PP} test further in \cref{sec:pp} and extend the investigation to also consider $\delta\mu_e\neq0$, showing how bias and scale-bias can be diagnosed from the failure of the \ac{PP} test.

Taking the posterior samples, we then calculate the weights as the ratio of the likelihood from the correct normal likelihood with standard deviation $\sigma_p$ to the erroneous distribution $\sigma_e$.
In \cref{fig:pp}, we then perform \ac{PP} tests for \ac{RS} and multinomial-\ac{IS} with and without Pareto smoothing.
For all four cases, the \ac{PP} test illustrates that the resampled posteriors are unbiased.

\begin{figure}
    \centering
    \includegraphics[width=\linewidth]{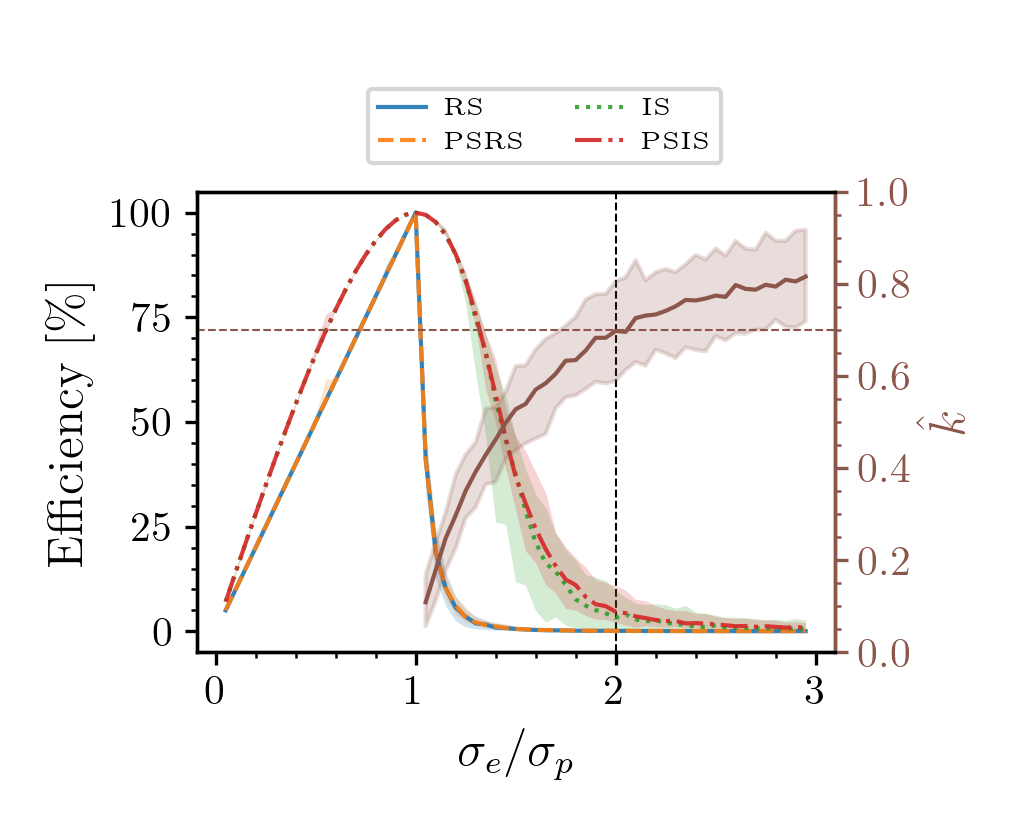}
    \caption{Measures of the efficiency of four resampling algorithms applied to concocted posterior distributions, described in \cref{sec:toymodel}.
    For \ac{RS} and \ac{PSRS}, we calculate the efficiency from the acceptance ratio of the resulting distribution as defined in \cref{eqn:eta_RS_def}.
    For \ac{IS} and \ac{PSIS}, we calculate the efficiency as the ratio of the \ac{ESS} to the number of original samples, as defined in \cref{eqn:eta_IS_def}.
    On the right-hand axes (brown solid curve), we plot the Pareto-smoothing diagnostic $\hat{k}$.
    For all quantities, we repeat the simulation study multiple times and then plot the median and 90\% interval across the simulated realisations.
    We provide a vertical dashed line at $\sigma_e/\sigma_p=2$, corresponding to the simulation studied in \cref{fig:pp}.
    We also provide a horizontal dashed line at $\hat{k}=0.7$ on the right-hand axis, a threshold above which the Pareto smoothing diagnostics suggest the estimates may be unstable or have a high degree of bias.
    }
    \label{fig:sigma_s_sigma_p}
\end{figure}

We then repeat the simulation study, varying $\sigma_p$ while keeping $\sigma_e$ fixed.
In \cref{fig:sigma_s_sigma_p}, we plot the efficiencies as calculated directly from \cref{eqn:eta_RS_def} and \cref{eqn:eta_IS_def}.
This allows us to study the performance of the different methods in practice (though, as discussed in \cref{sec:is}, the efficiency measures of \ac{RS} and \ac{IS} are not an apples-to-apples comparison).

For $\sigma_e/\sigma_p<1$, the generating distribution is broader than the target distribution (i.e. the concocted posterior is under-constrained).
This is the ideal case for resampling, and we find good efficiency with \ac{RS} behaving linearly.
Multinomial-\ac{IS} does yield greater efficiency (and hence will produce smoother histograms), but at the cost of repeated samples.
We do not find that Pareto smoothing has any impact in this regime.
This is further confirmed by the measurements of $\hat{k}$ (right-hand axes), which are less than $0$ for all $\sigma_s/\sigma_p<1$.

For $\sigma_e/\sigma_p>1$, the generating distribution is narrower than the target distribution (i.e. the concocted posterior is over-constrained).
In this case, the efficiency of \ac{RS} rapidly decays while \ac{IS} exhibits a more gradual decrease.
Pareto smoothing improves the performance of \ac{IS} by a few tens of per cent (and \ac{RS} to a lesser extent).
However, as shown on the right-hand axis, above $\sigma_e/\sigma_p > 2$, the median of the $\hat{k}$ distribution exceeds 0.7, a threshold above which the results can be biased \citep{vehtari2024pareto}.
For cases like this (and more extreme), where the generating distribution does not cover the target, it is generally advisable to use an alternative approach such as a re-analysis of the data, or sequential Monte Carlo as explored in \citet{Williams:2025szm}.

\subsection{Changing the waveform model for GW150914}
\label{sec:wf}
To illustrate the application of resampling to gravitational-wave observations, we take the GWTC-2.1 data released for the \ac{LVK} analysis of GW150914 \cite{LIGOScientific:2021usb, GWTC-2.1:PE} with the \IMRPhenomXPHM waveform model \citep{Pratten:2020ceb}.
We then reconstruct the likelihood, using the open data from \ac{GWOSC} \citep{GWTC-4-data}, and then calculate a new likelihood replacing the waveform with \IMRPhenomXP (a special case of \IMRPhenomXPHM which does not contain higher-order modes).
This is in contrast to how resampling would usually be used \citep[see, e.g][]{Payne:2019wmy}, but we provide it as a simple illustration of the techniques rather than a scientific result.
In particular, the \IMRPhenomXP primary pass posterior distribution is wider than the posterior found using the \IMRPhenomXPHM model.
This is expected, since the higher-order modes contain additional information about the source parameters that isn't captured by the dominant mode alone.

\begin{figure}
    \centering
    \includegraphics[]{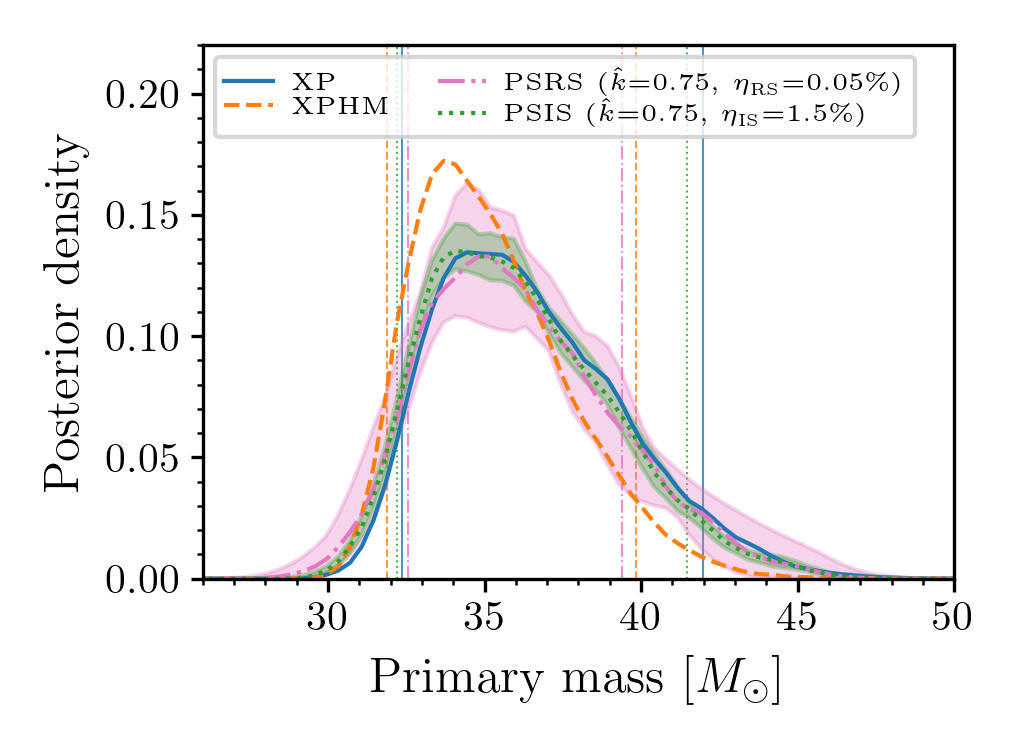}
    \caption{The posterior distribution (presented using a Gaussian \ac{KDE}) on the primary mass of GW150914 as presented in the GWTC-2.1 analysis using \IMRPhenomXPHM (orange dashed curve).
    We compare this with the posterior distributions for the \IMRPhenomXP waveform calculated from resampling the GWTC-2.1 results with \ac{PSRS} (pink dash-dotted curve) and \ac{PSIS} (green dotted curve).
    We also include a new re-analysis using identical settings to GWTC-2.1, but with the \IMRPhenomXP model (blue solid curve).
    For the resampled posteriors, we repeat the resampling multiple times and plot the median and 90\% intervals to illustrate the variance in the result.
    }
    \label{fig:GW150914-wf}
\end{figure}

In \cref{fig:GW150914-wf}, we plot the posterior distributions, comparing the original analysis (\IMRPhenomXPHM) with two sets of resampled posteriors using \ac{PSRS} and multinomial-\ac{PSIS} changing the waveform model.
To validate the resampling procedure, we also include a new re-analysis of GW150914, using an identical setup to the GWTC-2.1 analysis but changing only the waveform model.
The samples from this analysis agree within the resampling uncertainty with both \ac{PSRS} and \ac{PSIS}, demonstrating their capacity to correctly resample the posterior.
However, we do find that the Pareto smoothing diagnostic is greater than the threshold of $0.7$.
Finally, by showing the uncertainty in the inferred density (calculated by repeated resampling), we illustrate the key difference between \ac{IS} and \ac{RS}: the \ac{PSIS} interval is narrower than the \ac{PSRS} interval, indicating a reduction in the variance and a smoother resulting histogram. 
However, it should of course be recalled that the cost of using multinomial-\ac{PSIS} is repeated samples.

Finally, we also compute the Bayes factor between the \IMRPhenomXP and \IMRPhenomXPHM models (using the weights before Pareto smoothing) to be 
\begin{align}
    \ln \BF=\num{0.16}\pm{\num{0.03}}\,
\end{align}
which provides mild support in favour of the \IMRPhenomXP waveform model over \IMRPhenomXPHM.
This finding is consistent with \citet{Payne:2019wmy}: the log-Bayes factor is within 2 standard deviations of their value $\ln \BF =0.20$ (here we have changed the sign of the Bayes factor found in Table 1 of that work so that the definition of the Bayes factors are consistent).

\subsection{Changing the \ac{PSD} for GW150914}
\begin{figure}
    \centering
    \includegraphics[]{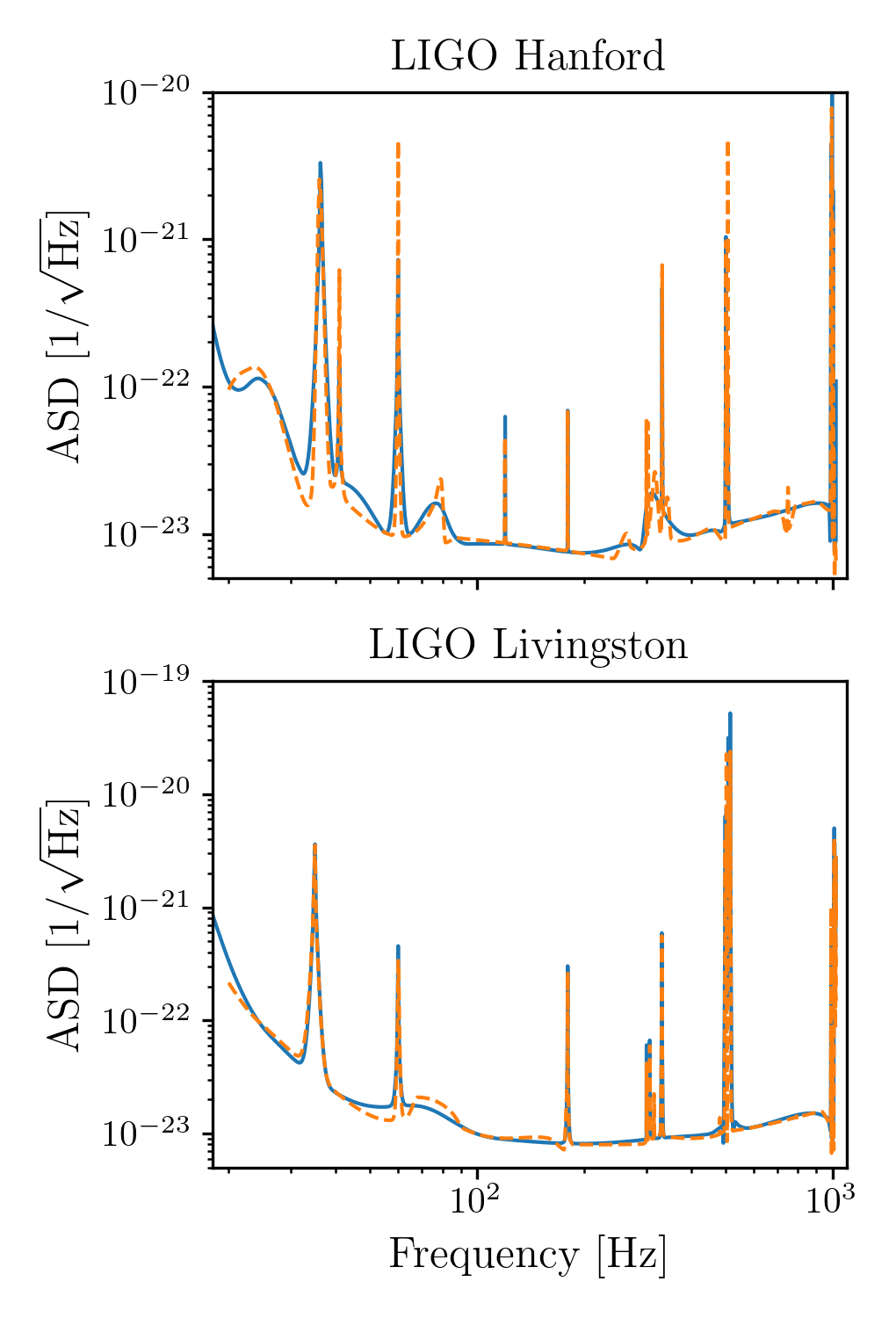}
    \caption{A comparison of the \BAYESWAVE \ac{ASD} estimates from the original GWTC-2.1 analysis (solid blue curves) with a new \ac{ASD} created using the upgraded version of \BAYESWAVE (orange dashed curves).}
    \label{fig:GW150914-psd}
\end{figure}

As another example of the application of resampling, we again consider the GWTC-2.1 re-analysis of GW150914.
For that analysis, \BAYESWAVE \citep{Littenberg:2014oda} was used to calculate a \ac{PSD}, using data surrounding the signal and modelling the noise properties with a parameterised model including cubic splines and Lorentzian lines.
However, the version of \BAYESWAVE used to produce \ac{LVK} parameter estimates has since been upgraded \citep{2024PhRvD.109f4040G}, for example, using Akima splines rather than cubic splines.
We use this upgraded version to calculate a new \ac{PSD} with settings similar to those used in the GWTC-4.0 analysis \citep{GWTC-4-methods}.
In \cref{fig:GW150914-psd}, we compare the \acsu{ASD} (\acl{ASD}: the square root of the \ac{PSD}) for the two \ac{LIGO} detectors that observed the event, illustrating that while the overall shape is consistent, there are small differences between the two.

\begin{figure}
    \centering
    \includegraphics[]{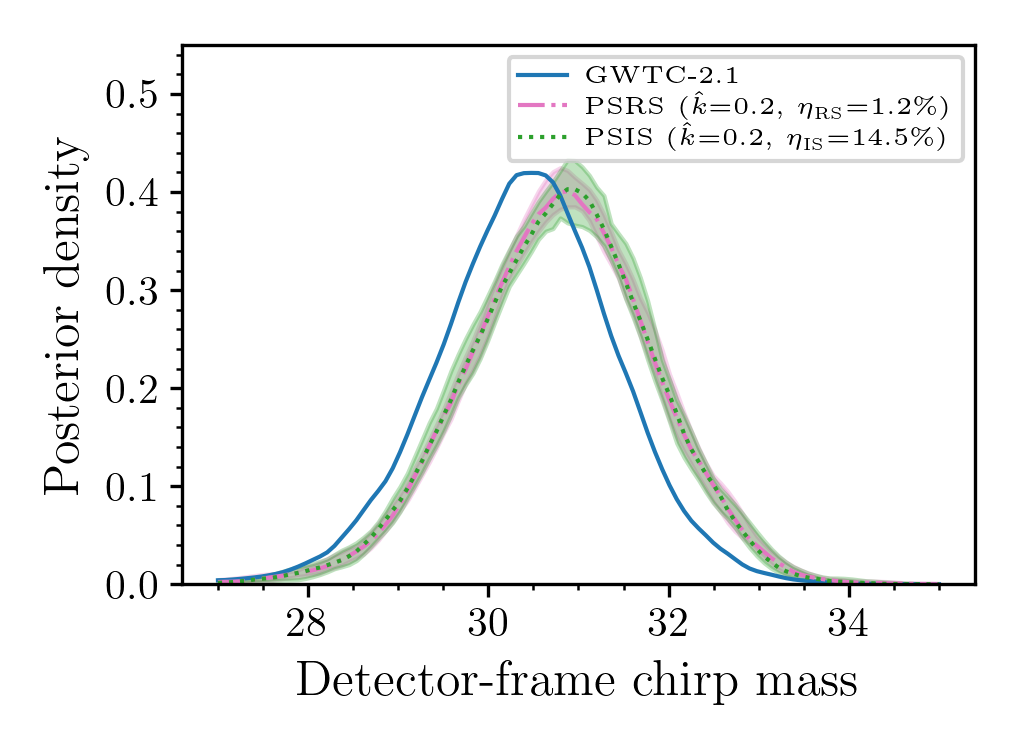}
    \caption{The redshifted (detector-frame) chirp mass posterior distribution (presented as a Gaussian \ac{KDE}) of GW150914 from the original GWTC-2.1 analysis (solid blue curve) and resampling using an updated \BAYESWAVE \ac{PSD} (see \cref{fig:GW150914-psd}) using \ac{PSRS} (pink dash-dotted curve) and \ac{PSIS} (green dotted curve).
    For the resampled distributions, we repeat the process multiple times and plot the median of the distribution and 90\% interval to illustrate the variance in the procedure.}
    \label{fig:GW150914-psd-mc}
\end{figure}

Resampling can provide an easy way to check the impact these differences may have on the scientific results.
However, we have to be careful in our choice of likelihood since the \ac{PSD} enters \cref{eqn:likelihood} in a normalisation term that is often ignored \citep[cf.][]{Thrane:2018qnx}.
This correction is not included in the \BILBY likelihood computation by default, and so we add it explicitly (see the data release for details).
We calculate the likelihood and likelihood ratio under the old and new \ac{PSD}
then using the difference in log-likelihoods, resample from the initial sample set.
In \cref{fig:GW150914-psd-mc}, we plot the distribution of the detector-frame (i.e. redshifted) chirp mass.
We select this parameter for visualisation as it is the best measured value.
We find that the change in \ac{PSD} results in a few-per-cent shift in the posterior, consistent with the findings of \citet{Biscoveanu:2020kat}.
We also find that the resampling is efficient, with a good Pareto-smoothing diagnostic and efficiency for both \ac{PSRS} and \ac{PSRS}.

Now turning to the Bayes factor, from \cref{eqn:likelihood-ratio}, we see that three different Bayes factors can be computed when changing from the ``old'' (GWTC-2.1) to our ``new'' \ac{PSD} (using the GWTC-4.0 settings).
First, the ratio of the signal evidence, second, the ratio of the noise evidence, and third, the ratio of the signal vs noise evidence.
We tabulate these in \cref{tab:bayes-factors}.
The first row, the Bayes factor of signal evidence, indicates a strong preference for the old \ac{PSD} over the new \ac{PSD}.
However, this comparison includes information about both how well the signal model explains the data and how well the \ac{PSD} explains the noise properties of the data.
We can separate these by considering the second row, the ratio of noise evidence, which indicates an even stronger preference for the old \ac{PSD}.
Finally, when we compute the ratio of the signal vs noise Bayes factors, we effectively ask ``Which PSD makes the signal stand out more strongly?'' and now we find evidence in favour of the new \ac{PSD}.
This suggests the new PSD, while assigning a lower overall probability to the data, provides a noise model against which the signal is more distinguishable.
We do not investigate this point further, but note that Bayes factors are not typically used as a means to determine the choice of \ac{PSD}: for details of the development of the \BAYESWAVE see \citet{2024PhRvD.109f4040G}.
Moreover, we highlight that the differences could be related to the settings we chose for the \ac{PSD} generation.

\begin{table}
    \centering
    \begin{tabular}{c|c}
        Bayes factor & Estimate \\\hline
         $\ln \left(\frac{\evidence(d| \model, P_{\rm new})}{\evidence(d| \model, P_{\rm old})}\right)$& $\num{-1684.86}\pm\num{0.02}$ \\
         $\ln \left(\frac{\evidence(d| N, P_{\rm new})}{\evidence(d| N, P_{\rm old})}\right)$& $\num{-1726.90}$ \\
         $\ln \left(\frac{\evidence(d| \model, P_{\rm new})}{\evidence(d| N, P_{\rm new})}
         \frac{\evidence(d| N, P_{\rm old})}{\evidence(d| \model, P_{\rm old})}\right)$& $\num{42.04}\pm{0.02}$ \\
    \end{tabular}
    \caption{
    Estimates of the three log Bayes factors that can be computed from the study changing the \ac{PSD} in the analysis of GW150914.
    For the first and last lines, the evidences are computed from \cref{eqn:BF} using different choices of the likelihood or likelihood ratio. Meanwhile, for the second row, the ratio of noise evidence, there is no uncertainty, as this is computed directly from the data and \ac{PSD}.
    Note that all three are computed directly from the likelihoods themselves, but the final row can also be computed as the difference in the first two rows.
    }
    \label{tab:bayes-factors}
\end{table}

%% file: content/5-discussion.tex
\section{Discussion}
\label{sec:discussion}

In this work, we have provided a comprehensive guide to reconstructing likelihoods and priors from gravitational-wave posterior samples and applying resampling techniques to generate new posterior distributions under modified assumptions.
This capability is fundamental to many analyses in gravitational-wave astronomy, from population studies to tests of general relativity, yet the practical implementation details have not been systematically documented until now.
Our analysis demonstrates that accurate reconstruction of the likelihood and prior is achievable when the computing environment and analysis configuration are properly matched. The residual differences we observe (\cref{fig:residuals}) are typically at the level of one part in a thousand or better for the likelihood, which we have shown is sufficiently accurate to have a negligible impact on resampling efficiency.
This level of precision requires careful attention to software versions, data processing details, and numerical implementation choices.
The hardware-dependent variations we identified highlight an important but often overlooked aspect of reproducible gravitational-wave analysis.
While exact byte-level reproduction requires identical hardware, the small random shifts introduced by different floating-point implementations are typically negligible for scientific applications, which should reassure researchers that resampling can be performed reliably across different computing environments.

A key limitation of our approach is the requirement that the original analysis was performed using \BILBY with \PESUMMARY packaging.
While this covers a significant fraction of \ac{LVK} analyses since GWTC-2.1, other analyses using \LALINFERENCE or \RIFT would require adapted approaches.
Additionally, our reconstruction methodology assumes that all necessary configuration information has been preserved in the data release, which may not always be the case for older analyses.

Our comparison of resampling techniques reveals important trade-offs that practitioners should consider.
\ac{RS} produces clean, equally-weighted samples but can be inefficient when the proposal and target distributions differ significantly.
Multinomial-\ac{IS} can yield smoother posterior estimates, but at the cost of introducing sample correlations through repeated samples in multinomial resampling.
For modest changes to analysis assumptions, both approaches perform well.
For more dramatic changes, \ac{IS} may be preferable to avoid significant variance in the resulting posterior samples, though a re-analysis would be better if computationally feasible.

We also introduce Pareto-smoothing, demonstrating that it can provide moderate improvements in efficiency and comes with an automated diagnostic which provides a useful indicator of when Pareto-smoothing can be trusted.

Our real-data examples demonstrate that some typical analysis modifications (waveform model changes, PSD updates) can fall within the regime where resampling is efficient and reliable.
However, we caution that more extreme modifications—such as changing the waveform family, changing from a precessing to non-precessing waveform or dramatically altering the frequency band—may fail altogether or require careful validation of the resampling assumptions.
The GW150914 examples illustrate that even seemingly minor changes, such as waveform model updates or improved PSD estimation algorithms, can have measurable impacts on inferred parameters, making resampling a valuable tool for systematic uncertainty assessment.

The methodology presented here enables several important classes of scientific analysis that would otherwise require computationally expensive re-analysis.
Therefore, resampling can allow researchers without access to high-performance computing resources to explore ``what if'' questions that would otherwise require full re-analysis.

Based on our analysis, we recommend several best practices for gravitational-wave posterior resampling.
When possible, researchers should match the \CONDA environment used in the original analysis, using the environment specifications provided in Table 1 and our accompanying data release.
For method selection, rejection sampling is preferred for modest analysis changes where clean, equally-weighted samples are needed, while importance sampling with Pareto-smoothing should be used for larger changes or when smooth posterior estimates are critical, carefully monitoring the $\hat{k}$ diagnostic.
When possible, we recommend validation against a full re-analysis for at least one representative event.
Calculating and reporting resampling efficiency helps readers assess the reliability of results, as low efficiency may indicate that the analysis modification is too extreme for reliable resampling.

We hope this guide provides a useful resource for the gravitational-wave community and enables new scientific investigations that leverage the wealth of information contained in the LVK data releases. The combination of theoretical understanding, practical implementation details, and working code examples should lower the barrier for researchers to incorporate resampling techniques into their analyses, ultimately leading to more robust and comprehensive studies of the gravitational-wave universe.

%% file: content/6-appendix.tex
\section{Probability-probability tests}
\label{sec:pp}

\begin{figure*}
    \centering
    \includegraphics[]{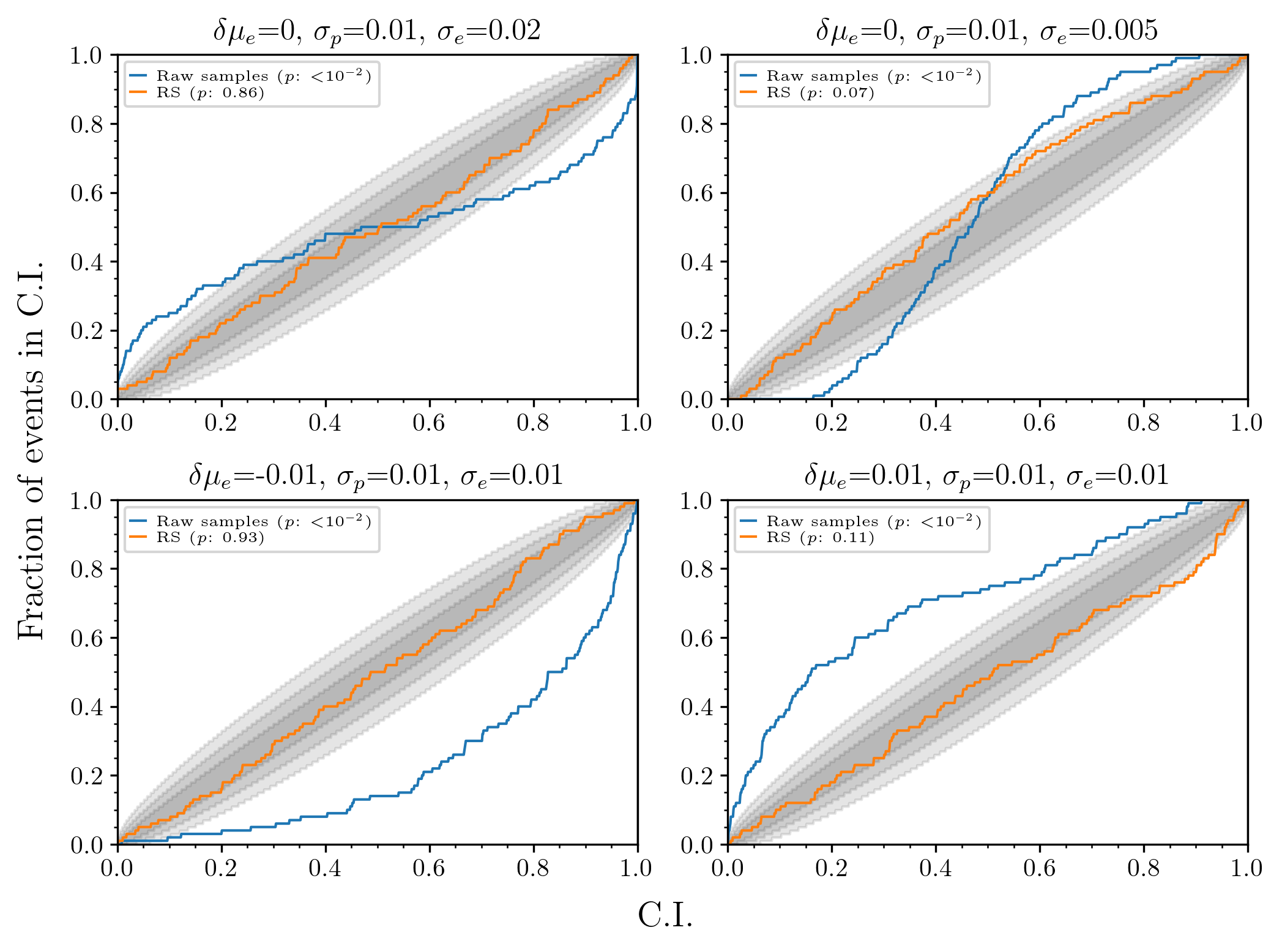}
    \caption{A set of four \ac{PP} tests showing the four failure modes of the test.
    In blue, we plot the biased tests, and in orange, we plot the corrected posteriors.
    For each plot, we provide the parameters of the simulation in the title.
    Further details of the plots can be found in the notebooks as part of the data release.
    }
    \label{fig:pp-extended}
\end{figure*}

The \ac{PP} test is a standard diagnostic tool used to assess the reliability of Bayesian posterior distributions in gravitational-wave parameter estimation \citep{Veitch:2014wba, Romero-Shaw:2020owr}.
The test compares the cumulative distribution of confidence intervals from a set of simulated injections against the uniform distribution expected for unbiased posteriors. 
We follow the \BILBY implementation of the \ac{PP} test and provide a simplified version in the data release notebooks.
Specifically, we take the result from injection studies and compute the fraction of injected parameter values that fall below various quantile levels of their corresponding posterior distributions.
For each parameter of interest, the \ac{PP} test calculates the cumulative probability $P(\theta < \theta_{\rm sim}$) where $\theta_{\rm sim}$ is the simulated value and the probability is computed using the recovered posterior samples.
If the posteriors are unbiased and properly calibrated, these $p$-values should be uniformly distributed between 0 and 1.
The \ac{PP} plot displays the empirical cumulative distribution function of these $p$-values against the theoretical uniform distribution, with deviations from the diagonal indicating systematic biases in the parameter estimation.
We include a Kolmogorov-Smirnov test to compare the empirical distribution of $p$-values against the uniform distribution, providing a $p$-value that indicates the probability of observing such deviations by chance.
In \cref{fig:pp-extended}, we also include confidence bands around the diagonal using the beta distribution to account for finite sample size effects.
The width of these bands depends on the number of injections, with larger injection campaigns providing tighter constraints on systematic biases.

In the main text, Section~\ref{sec:example}, we use the \ac{PP} test to demonstrate that the resampling methods can unbias a posterior given a corrected posterior density.
The specific case given considers over- and under-constrained posteriors generated by varying the ratio of $\sigma_s$ (the standard deviation of the simulated measurement error)  to $\sigma_p$ (the standard deviation of the simulated posteriors).
Specifically, in \cref{fig:pp}, we consider the case when the posteriors are too narrow $\sigma_s/\sigma_p=2$.
This results in a characteristic  `S' shape.
This is one possible characteristic failure mode of the \ac{PP} plot.
In \cref{fig:pp-extended}, we extend this to study the other three failure modes: a case where the posterior is under-constrained (producing an inversion of the `S'); biased positively (producing a bow-like deviation above the diagonal), and negatively (producing a bow-like deviation below the diagonal).
We include these figures to help users diagnose bias from resampling (or from direct sampling) approaches.

We note that this behaviour is specific to \ac{PP} tests, which use quantile levels to calculate the fraction of injected parameters that fall outside the posterior at a given level.
If instead, the highest-posterior density is used, the behaviour will differ.

